\documentclass[%
 reprint,
 amsmath,amssymb,
 aps,
]{revtex4-2}

\usepackage{graphicx}
\usepackage{dcolumn}
\usepackage{bm}
\usepackage{multirow}
\usepackage{nomencl}
\usepackage{hyperref}

\usepackage{caption}
\usepackage{subcaption}
\usepackage{gensymb}
\usepackage{xurl}

\usepackage{threeparttable}

\begin{document}

\preprint{APS/123-QED}

\title{Investment-based optimisation of energy storage design parameters in a grid-connected hybrid renewable energy system}

\author{Sleiman Farah}
\email{sleiman.farah@mpe.au.dk}
\affiliation{%
 Department of Mechanical and Production Engineering, Aarhus University, Denmark}%

\author{Gorm Bruun Andresen}%
\affiliation{%
 Department of Mechanical and Production Engineering, Aarhus University, Denmark}%




\date{\today}

\begin{abstract}
Grid-connected hybrid renewable power systems with energy storage can reduce the intermittency of renewable power supply. However, emerging energy storage technologies need improvement to compete with lithium-ion batteries and reduce the cost of energy. Identifying and optimizing the most valuable improvement path of these technologies is challenging due to the non-linearity of the energy system model when considering parameters as independent variables. To overcome this, a novel investment-based optimisation method is proposed. The method involves linear optimization of the hybrid renewable energy system and subsequent investment optimization, accounting for diminishing improvements per investment. The results from applying the investment-based optimisation to thermal energy, pumped thermal energy, molten salt, and adiabatic compressed air energy storage technologies, show that improving the discharge efficiency is the most valuable for all technologies. The second most important parameters are the costs of discharge capacity and energy storage capacity, and the least important parameters are the charge capacity cost and charge efficiency . The study provides detailed improvement pathways for each technology under various operational conditions, assisting developers in resource allocation. Overall, the investment-based optimization method and findings contribute to enhancing the competitiveness of emerging energy storage technologies and reducing reliance on batteries in renewable energy systems.
\end{abstract}

\maketitle


\section{\label{S:Introduction}Introduction}

Decarbonising the electricity grid by transitioning away from fossil fuels towards renewable energy sources is essential to reduce greenhouse gas emissions and combat climate change~\cite{renewables2022}. However, the integration of intermittent renewable power sources, such as solar and wind, increases the difficulty of managing the electricity grid and maintaining the balance of electricity supply and demand~\cite{Maintaining2017}.

To alleviate the adverse effects of renewable energy integration into the grid, some recent tenders of grid-connected renewable energy systems now mandate that annual net energy supplied to the grid should exceed a minimum threshold specified by the grid-connection capacity factor. Typically, the requirement exceeds the capacity factor of local renewable resources~\cite{Dibyanshu2020}. For instance, a renewable energy system with 100~MW grid-connection capacity and 50\% capacity factor should supply at least 438~GWh net renewable electricity to the grid annually. To achieve the required energy supply for relatively low capacity factors, renewable energy systems are likely to consist of either, solar, wind, or a combination of both electricity sources depending on the availability of natural sources. However, for high capacity factors of the grid-connection, the capacities of renewable energy sources increase significantly, adversely affecting the economic viability of the system.

Therefore, for high grid-connection capacity factors, i.e., high renewable energy penetration into the grid, energy storage becomes important to compensate for the intermittency of renewable electricity sources and for reducing the system capacities. Energy storage systems can store excess energy generated by renewable sources during periods of high generation, and supply the previously stored energy into the grid when the renewable electricity generation is low. This ability of energy storage systems reduces the undesirable intermittency effect and enables more efficient use of renewable energy sources.

Several emerging energy storage technologies can fulfill the role of reducing the intermittency of renewable energy sources. However, emerging technologies are typically competing with lithium-ion batteries that exhibit excellent performance characteristics to fulfill the same role~\cite{schmidt2017future, KECK2019647, LENZEN2016553}. Lithium-ion batteries may dominate the energy storage market especially that they are suitable for multiple applications that can increase their market penetration~\cite{schmidt2017future}.

Therefore, from the perspective of energy storage developers, identifying and improving important characteristics of emerging technologies is critical to be able to become competitive with lithium-ion batteries. Given limited resources, however, should the development efforts be on improving technical characteristics or on reducing capacity costs for specific energy storage technologies? Energy storage developers are most likely interested in identifying the optimal distribution of resources or investments to improve the characteristics of specific energy storage technologies. 

The need for energy storage for renewable energy integration into the electricity grid is highlighted in several studies~\cite{electronics8070729, VICTORIA2019111977, BROWN2018720, sepulveda2021design}. The role of energy storage in the feasibility of renewable energy integration into the European electricity network was analysed for different scenarios~\cite{electronics8070729}. The average storage fraction was calculated as a function of variable renewable energy penetration using a polynomial regression model. The overall conclusion was that increasing the use of energy storage by energy storage market development and regulations is essential for successful renewable energy integration~\cite{electronics8070729}.

The role of energy storage in the European energy network was also modelled considering coupling of electricity, transport and heating sectors for various CO$_2$ reduction constraints~\cite{VICTORIA2019111977}. The network was modelled in PyPSA-Eur-Sec-30~\cite{BROWN2018720}, considering hourly resolution and 30 European countries. The model calculated the optimal capacities and optimal power flows within the energy network. The results showed that sector coupling reduces the need for large energy storage capacities which become necessary at high CO$_2$ reductions~\cite{VICTORIA2019111977}.

The aforementioned studies reveal the importance of energy storage systems especially with high penetration of renewable energy. However, these studies do not investigate the effect of energy storage parameters at the technology level, i.e., they do not analyse the effect of design parameters of energy storage technologies. Such analysis was conducted considering two energy systems in the United States of America~\cite{sepulveda2021design}. Discrete combinations of efficiency and cost values for existing and potential future long-duration energy storage technologies were utilised in an electricity system capacity expansion model. The analysed parameter combinations achieved different system values defined as reductions in the total electricity system. These reductions were utilised in a regression model to identify the order of significance of the storage parameters. The energy capacity cost was the most important parameter followed by the discharge efficiency and the discharge capacity cost. The least two important parameters were the charge efficiency and the charge efficiency cost respectively~\cite{sepulveda2021design}.

In a similar study of the European energy network~\cite{PRXEnergy.2.023006}, the parameters of energy storage were analysed considering three scenarios: electricity only, electricity with heating and land transport, and fully-coupled energy sectors. The aim was to analyse the potential of emerging long-duration storage technologies with large implementation of renewable power sources to limit CO$_2$ emissions to 5\% of those of 1990. A large number of storage designs was generated for the analysis based on a discrete parametric space. The generated storage designs were then utilised in the energy system optimisation to identify the storage designs that could have a significant energy capacity deployment ($\ge 2~TWh)$ in the optimal energy system configuration. The results showed that storage designs with relatively high energy capacity cost or low discharge efficiency could not satisfy the energy capacity constraint. These results indicated that the energy capacity cost and the discharge efficiency represent the two most important parameters of the storage design. The results were also supported by the regression coefficient of a multi-variable linear regression model. The regression model had the energy storage capacity deployment as the dependent variable and the energy storage parameters as the independent variables. The energy capacity cost and the discharge efficiency had the largest normalised coefficient values~\cite{PRXEnergy.2.023006}. The charge capacity cost was the next important parameter after the energy capacity cost and the discharge efficiency. The least important parameters were the charge efficiency and the discharge capacity cost.

The results presented in both~\cite{sepulveda2021design} and~\cite{PRXEnergy.2.023006} provide insightful information regarding the parameters of emerging long-duration storage technologies for high penetration of renewable energy. A comparison between the order of parameters importance reveals that the discharge efficiency is an important parameter. However, the comparison is inconclusive with regards to the remaining parameters. The results derived from the multi-variable linear regression models to determine the order of importance of storage parameters are also susceptible to bias based on the chosen sampling space of the storage parameters. The multi-variable linear regression models reveal only the general trend in the analysed data; they do not necessarily identify the accurate order of importance of energy storage parameters for specific technologies, particularly when the regression adjusted $R^2$-value is low.

Further, while large renewable energy system models that incorporate sector-coupling, as those in~\cite{sepulveda2021design, PRXEnergy.2.023006}, offer a more holistic view of the energy system, they inherently exhibit greater uncertainties that could impact the results. Large-scale models also include the implicit assumption that a single entity controls the allocation of renewable power sources and energy storage systems to different location that may spread over several states or countries. However, renewable energy systems are typically built based on localised system optimisation and decision-making processes. In other words, renewable energy systems are built based on their techno-economic assessment, regardless whether the system being built is aligned with the configuration identified from the optimisation of large renewable energy systems that cross multiple jurisdictions.    

These factors emphasise the need to consider a complementary approach when evaluating the order of importance of storage parameters in an energy system. Therefore, this research presents an investment-based optimisation method of energy storage parameters in a grid-connected hybrid renewable energy system. The investments are allocated optimally to improve the energy storage parameters with the objective of minimising the Levelised Cost Of Energy ($LCOE$). The order of importance of energy storage parameters is determined by their corresponding optimal order of investments allocations. The investment-based optimisation method also allows focusing on specific emerging energy storage technologies instead of providing a single order of importance that should relatively represent all technologies. The optimisation method can be applied to large and complex energy systems, as those described in~\cite{HENNI2021116597, GEABERMUDEZ2021116685, PRXEnergy.2.023006, sepulveda2021design}. However, a simpler energy system is adopted in this research to present the optimisation method and to reduce the uncertainties associated with large and complex energy systems; the adopted renewable energy system is likely to resemble the structure of new grid-connected renewable energy systems. The adopted energy system is also localised and can be analysed as a single business case for renewable power and energy storage developers. 

In energy system modelling, the optimisation of energy storage parameters is computationally challenging; considering the parameters as independent optimisation variables breaks the linearity of the energy system model and complicates the system optimisation. The novelty in this research is the optimisation method that maintains the optimisation linearity by adopting a two-step optimisation. The first step is a typical capacity and power dispatch optimisation of the energy system, and the second is a linear investment optimisation that considers decaying improvement of the storage parameters per investment. The second optimisation step is fundamentally different from the discrete sampling and linear regression presented in~\cite{sepulveda2021design, PRXEnergy.2.023006}. The method also overcomes the limitations of typical sensitivity analyses of energy storage parameters to identify their importance. The method considers the improvement potentials of energy storage parameters that are ignored in sensitivity analyses.

The description and optimisation model of the hybrid renewable energy system are presented in Section~\ref{S:Hybrid renewable system}, and the optimisation method of the energy storage parameters is presented in Section~\ref{S:Investment-based optimisation of energy storage parameters}. Section~\ref{S:Optimal improvements of emerging energy storage technologies} presents the details of the case study and the corresponding results including the order of importance of storage parameters. Section~\ref{S:Discussion} discusses the results from Section~\ref{S:Optimal improvements of emerging energy storage technologies} combined with the results for different capacity factors for solar, wind and grid connection. Section~\ref{S:Discussion} also presents a comparison with results from the literature and the limitations of the developed method and results. Section~\ref{S:Summary and conclusions} presents the summary and conclusions of the research, as well as potential future research related to analysing the order of importance of energy storage parameters.

\section{\label{S:Methodology}Methodology}

This section presents the hybrid renewable energy system and the corresponding optimisation model that calculates the optimal capacities and power flows with the objective of minimising the $LCOE$. This section also presents the investment-based optimisation method that calculates the optimal allocation of a series of investments for improving energy storage parameters. This method also considers the objective of minimising the $LCOE$.

\subsection{\label{S:Hybrid renewable system}Hybrid renewable system}

\subsubsection{System description}

\begin{figure}
    \includegraphics[trim={5.0cm 5.0cm 5.0cm 5.0cm}, clip, width=0.45\textwidth]{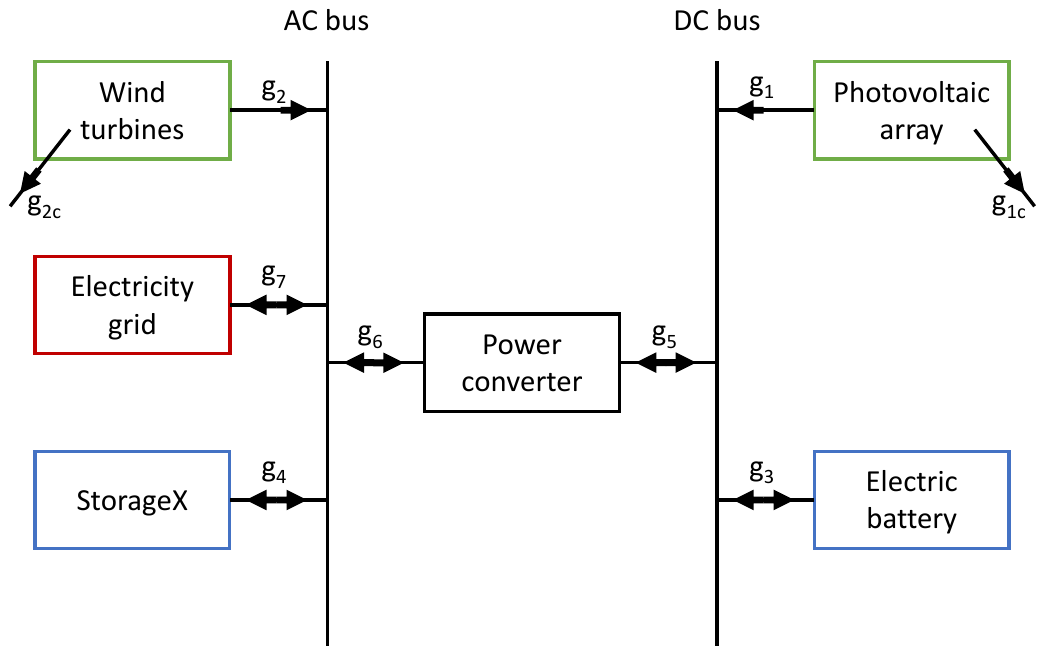}
    \caption{Schematic of hybrid renewable energy system.}
    \label{F:HPSS}
\end{figure}

The hybrid renewable energy system consists of two power sources, solar and wind, and two energy storage systems, battery and storageX (see Fig.~\ref{F:HPSS}). StorageX refers to a generic energy storage technology, such as Thermal Energy Storage (TES). The solar and battery systems are connected to the direct current (DC) bus, whereas the wind and storageX systems are connected to the alternating current (AC) bus. Electric power can flow through the power converter from the DC bus to the AC bus (inverter) and vice versa (rectifier). The hybrid renewable system can export and import electricity from the grid which is connected to the AC bus.

A typical optimisation model of the hybrid renewable energy system is presented in Section~\ref{S:System model}. The optimisation model allows calculating the minimum ($LCOE$) needed for the optimisation of energy storage parameters in Section~\ref{S:Investment-based optimisation of energy storage parameters}. The optimisation model details of the hybrid renewable energy system are presented for completeness and clarity; readers familiar with optimisation of such systems can continue from Section~\ref{S:Investment-based optimisation of energy storage parameters}.

\subsubsection{\label{S:System model}System model}
The system model consists of the energy and power balances, as well as  constraints of components and power flows of the hybrid system. The model is developed in Python~\cite{van1995python}, and the optimisation of capacities and power flows is resolved using Gurobi~\cite{gurobi}.

\begin{itemize}
   \item{DC bus power balance}
\end{itemize}
\begin{equation}
\label{E:DCBalance}
    g_{1,t} + g_{3,t} = g_{5,t} \le G_5/\eta_{i}
\end{equation}
\nomenclature{$g_{1,t}~(MW)$}{Solar power supplied to the DC bus}
\nomenclature{$g_{3,t}~(MW)$}{Battery discharge ($+$) or charge ($-$) power}
\nomenclature{$g_{5,t}~(MW)$}{Power from ($+$) or to ($-$) the DC bus}
\nomenclature{$G_5~(MW)$}{Power output capacity of the power converter (inverter)}

where $g_{1,t}$, $g_{3,t}$, $g_{5,t}$, $G_5$, and $\eta_{i}$ represent the solar power supplied to the DC bus, battery power, converter power (DC side), converter (inverter) capacity, and converter (inverter) efficiency respectively. The subscript $t$ represents hour of the year.

\begin{itemize}
    \item Solar power output considering curtailment
\end{itemize}
\begin{equation}
\label{E:SolarBalance}
    g_{1,t} = G_1 \times CF_{G_1,t} - g_{1c,t} \ge 0
\end{equation}
\nomenclature{$G_{1}~(MW)$}{Solar power capacity}
\nomenclature{$CF_{G_1,t}~$(-)}{Capacity factor of solar power at time t}
\nomenclature{$g_{1c,t}~(MW)$}{Curtailed solar power}

where $G_{1}$, $CF_{G_1,t}$, and $g_{1c,t}$ represent the solar capacity, solar capacity factor, and curtailed solar power respectively.
\begin{itemize}
    \item AC bus power balance 
\end{itemize}
\begin{equation}
    g_{2,t} + g_{4,t} - g_{7,t} = g_{6,t} \le G_6/\eta_{r}
\end{equation}
\nomenclature{$g_{2,t}~(MW)$}{Wind power supplied to the AC bus}
\nomenclature{$g_{4,t}~(MW)$}{Thermo-mechanical discharge ($+$) or charge ($-$) power}
\nomenclature{$g_{6,t}~(MW)$}{Power from ($+$) or to ($-$) the AC bus}
\nomenclature{$g_{7,t}~(MW)$}{Power exported ($+$) or imported ($-$) from the grid}
\nomenclature{$G_6~(MW)$}{Power output capacity of the power converter (rectifier)}

where $g_{2,t}$, $g_{4,t}$, $g_{6,t}$, $g_{7,t}$, $G_6$, and $\eta_{r}$ represent the wind power supplied to the AC bus, StorageX power, converter power (AC side), grid power, converter (rectifier) capacity, and converter (rectifier) efficiency respectively.

\begin{itemize}
    \item Wind power output considering curtailment
\end{itemize}
\begin{equation}
\label{E:WindBalance}
    g_{2,t} = G_2 \times CF_{G_2,t} - g_{2c,t} \ge 0
\end{equation}
\nomenclature{$G_{2}~(MW)$}{Wind power capacity}
\nomenclature{$CF_{G_2,t}~$(-)}{Capacity factor of solar power at time t}
\nomenclature{$g_{2c,t}~(MW)$}{Curtailed wind power}

where $G_{2}$, $CF_{G_2,t}~$, and $g_{2c,t}$ represent the wind capacity, wind capacity factor and curtailed wind power respectively.

\begin{itemize}
    \item Power limit of the grid connection
\end{itemize}
\begin{equation}
    G_{7,import} \le g_{7,t} \le G_{7,export}
\end{equation}

where $G_{7,import}$ and $G_{7,export}$ are the grid connection capacity of electricity import and export to the grid respectively.

\begin{itemize}
    \item Converter power balance (power flow from DC to AC)
\end{itemize}
\begin{equation}
    g_{6,t} = -g_{5,t} \times \eta_{i}
\end{equation}
\nomenclature{$\eta_{i}~$(-)}{Efficiency of converting DC to AC power}

\begin{itemize}
    \item Converter power balance (power flow from AC to DC)
\end{itemize}
\begin{equation}
    g_{5,t} = -g_{6,t} \times \eta_{r}
\end{equation}
\nomenclature{$\eta_{r}~$(-)}{Efficiency of converting AC to DC power}

\begin{itemize}
    \item Battery energy balance (discharge)
\end{itemize}
\begin{equation}
   E_{3,min} \le e_{3,t+1} = e_{3,t} \times (1 - L_3\times \Delta t/24) - \frac{g_{3,t} \times \Delta t}{\eta_{3,d}} 
\end{equation}
\nomenclature{$E_{3,min}$~(MWh)}{Minimum threshold of the battery energy content}
\nomenclature{$e_{3,t}$~(MWh)}{Battery energy content}
\nomenclature{$\eta_{3,d}~$(\%)}{Battery discharge efficiency}

where $E_{3,min}$, $e_{3,t}$, $e_{3,t+1}$, $L_3$, $\Delta t$, and $\eta_{3,d}$ represent the minimum energy content, energy content at time $t$, energy content at time ($t+1$), daily energy loss percentage, time step in hours, and discharge efficiency respectively.

\begin{itemize}
    \item Battery minimum energy content
\end{itemize}
\begin{equation}
    E_{3,min} = E_{3,cap} \times SOC_{3,min}
\end{equation}

where $E_{3,cap}$ and $SOC_{3,min}$ represent the energy storage capacity and the minimum state of charge respectively.

\begin{itemize}
    \item Battery energy balance (charge)
\end{itemize}
\begin{equation}
    e_{3,t+1} = e_{3,t} \times (1 - L_3\times \Delta t/24) - g_{3,t} \times \Delta t \times{\eta_{3,c}} \le E_{3,cap}
\end{equation}
\nomenclature{$\eta_{3,c}~$(\%)}{Battery charge efficiency}
\nomenclature{$E_{3,cap}$~(MWh)}{Battery energy storage capacity}

where $\eta_{3,c}$ represents the charge efficiency.

\begin{itemize}
    \item Battery power limits
\end{itemize}
\begin{equation}
    G_{3,charge} \le g_{3,t} \le G_{3,discharge}
\end{equation}

where $G_{3,charge}$ and $G_{3,discharge}$ represent the charge and discharge capacities respectively.

\begin{itemize}
    \item StorageX energy content balance (discharge)
\end{itemize}
\begin{equation}
    E_{4,min} \le e_{4,t+1} = e_{4,t} \times (1 - L_4\times \Delta t/24) - \frac{g_{4,t} \times \Delta t}{\eta_{4,d}}
\end{equation}
\nomenclature{$ECT_{min}$~(MWh)}{Minimum threshold of the thermo-mechanical energy storage content}
\nomenclature{$ECT_t$~(MWh)}{Energy content of StorageX}
\nomenclature{$\eta_{4,d}~$(\%)}{StorageX discharge efficiency}
\nomenclature{$L~$(-)}{Percentage of StorageX energy content lost to the environment each day}

where $E_{4,min}$, $e_{4,t}$, $e_{4,t+1}$, $L_4$, and $\eta_{4,d}$ represent the minimum energy content, energy content at time t, energy content at time ($t+1$), daily energy loss percentage, and discharge efficiency respectively.

\begin{itemize}
    \item StorageX minimum energy content
\end{itemize}
\begin{equation}
    E_{4,min} = E_{4,cap} \times SOC_{4,min} 
\end{equation}

where $E_{4,cap}$ and $SOC_{4,min}$ represent the energy storage capacity and the minimum state of charge respectively.

\begin{itemize}
    \item StorageX energy content balance (charge)
\end{itemize}
\begin{equation}
\label{E:TMBalanceCharge}
    e_{4,t+1} = e_{4,t} \times (1 - L_4\times \Delta t/24) - g_{4,t} \times \Delta t \times{\eta_{4,c}} \le E_{4,cap}
\end{equation}
\nomenclature{$\eta_{4,c}~$(\%)}{StorageX charge efficiency}
\nomenclature{$ECT_{cap}$~(MWh)}{StorageX energy storage capacity}

where $\eta_{4,c}$ represents the charge efficiency.

\begin{itemize}
    \item StorageX power limits
\end{itemize}
\begin{equation}
    G_{4,charge} \le g_{4,t} \le G_{4,discharge}
\end{equation}

where $G_{4,charge}$ and $G_{4,discharge}$ represent the charge and discharge capacities respectively.

To determine the optimal capacities and power flows, the system model is optimised considering that the annual net energy supply to the grid should satisfy a minimum threshold as expressed in Eq.~(\ref{E:ConstraintAnnualEnergy}).

\begin{equation}
    \sum_{t=1}^{8760/\Delta t} g_{7,t} \times \Delta t\ge 8760 \times G_{7,export} \times CF_{G_7}
    \label{E:ConstraintAnnualEnergy}
\end{equation}

\nomenclature{$\Delta t~(h)$}		{Time step}
\nomenclature{$t$~(-)}{Discrete time variable}
\nomenclature{$CF_{G_7}$~(-)}{Capacity factor of the grid connection}

where $CF_{G_7}$ represent the capacity factor of the grid connection.

The objective of the optimisation is to minimise the $LCOE$ which can be achieved by the minimisation of the system cost (SC) as shown in Eq.~(\ref{E:ObjFun})

\begin{equation}
    minimise~SC = \sum_{k=1}^{5} c^a_k\times G_k
    \label{E:ObjFun}
\end{equation}

where $c^a_k$ and $G_k$ are the annualised cost per unit capacity of technology $k$ and the capacity of technology $k$ respectively. The $c^a_k$ is calculated as a function of the annual fixed cost per unit capacity ($f_k$), the investment cost per unit capacity ($c_k$), the discount rate ($d$), and the technology lifetime ($T_k$) as shown in Eq.~(\ref{E:AC})
\begin{equation}
    c^a_k = f_k + \frac{ c_k \times d}{1-\left ( 1+d\right)^{-T_k}}
    \label{E:AC}
\end{equation}
The $LCOE$ is calculated as shown in Eq.~(\ref{E:LCOE})
\begin{equation}
    LCOE = \frac{SC^*}{8760 \times G_{7,export} \times CF_{G_7}}
    \label{E:LCOE}
\end{equation}
where $SC^*$ is the minimum system cost.

The optimisation model does not guarantee nor enforce any of the system components to be included in the optimal configuration of the hybrid renewable system. For example, the optimal energy system may include only one power source and either the battery or StorageX. When StorageX is excluded from the optimal system configuration, the $LCOE$ becomes independent or insensitive to StorageX parameters. In such case, the $LCOE$ calculation is pointless for the optimisation of StorageX parameters presented in Section~\ref{S:Investment-based optimisation of energy storage parameters}. Therefore, to ensure that StorageX is included in the optimal system configuration, and that the $LCOE$ is sensitive to StorageX parameters, an additional constraint is added to the optimisation model as shown in Eq.~(\ref{E:ConstraintAnnualEnergy_1})

\begin{equation}
    \sum_{t}^{8760} g_{4,t} \times \Delta t\ge 0.01 \times 8760 \times G_{7,export} \times CF_{G_7}
    \label{E:ConstraintAnnualEnergy_1}
\end{equation}

Eq.~(\ref{E:ConstraintAnnualEnergy_1}) ensures that a small percentage (1\%) of the energy supplied to the grid is from StorageX. This constraint is activated only when initial optimisation excludes StorageX from the optimal system configuration.

\subsection{\label{S:Investment-based optimisation of energy storage parameters}Investment based optimisation of energy storage parameters}

The system model allows minimising the $LCOE$ for a designated storage technology. Nevertheless, the system model does not provide insights into the optimisation of the parameters of a specific storage technology with the same objective of $LCOE$ minimisation. Hence, the optimisation of the storage parameters is achieved by the investment-based optimisation model presented in this section. The method is suitable for all storage parameters, however, for simplicity and based on results in the literature~\cite{PRXEnergy.2.023006}, the most significant parameters $\eta_c$, $\eta_d$, $c_c$, $c_d$, and $c_s$ are considered in the investment based optimisation model. In this method, the $LCOE$ is represented as a function of the storage parameters as shown in Eq.~(\ref{E:LCOEfunction}) 

\begin{equation}
\label{E:LCOEfunction}
\begin{aligned}
LCOE = g(\Vec{p})~~~~~ \Vec{p} & = (p_i~|~i=1, 2, ..., 5)\\
& = (\eta_c, \eta_d, c_c, c_d, c_s)   
\end{aligned}
\end{equation}

where the storage parameters are represented by vector $\Vec{p}$ to simplify the mathematical notations utilised in the subsequent calculations.

Considering that an investment $(I)$ is available, the aim is to calculate the optimal distribution of $I$, i.e., calculate the optimal sub-investments $(I_i)$ associated with improving parameter $p_i$. Hence, investment $I$ and the change of $LCOE$ with respect to $I$ can be represented as per Equations~\ref{E:I} and \ref{E:dLCOE/dI} respectively

\begin{equation}
\label{E:I}
    I = \sum_{i=1}^5 I_i
\end{equation}
\begin{equation}
\label{E:dLCOE/dI}
\frac{\mathrm{d}LCOE}{\mathrm{d}I} = \sum_{i=1}^5 \frac{\partial g}{\partial p_i}\frac{\mathrm{d}p_i}{\mathrm{d}I_i}\frac{\mathrm{d}I_i}{\mathrm{d}I}
\end{equation}

where $\frac{\partial g}{\partial p_i}$ is the $LCOE$ improvement per parameter improvement, $\frac{\mathrm{d}p_i}{\mathrm{d}I_i}$ is the parameter improvement per investment ($\mathrm{d}I_i$), and $\frac{\mathrm{d}I_i}{\mathrm{d}I}$ is the investment fraction allocated to improving parameter $p_i$.\\
The term $\frac{\partial g}{\partial p_i}$ can be estimated numerically as shown in Eq.~(\ref{E:Gradient})

\begin{equation}
\label{E:Gradient}
\frac{\partial g}{\partial p_i} \Big|_{\Vec{p}} \approx \frac{LCOE|_{p_i + \Delta p_i} - LCOE|_{p_i}}{\Delta p_i} 
\end{equation}

where the $LCOE$ can be calculated for a given $\Vec{p}$ from the optimisation of the energy system model, and $\Delta p_i =0.0001$ is a small change of parameter $p_i$.

\begin{figure}
    \includegraphics[clip, trim=0cm 0cm 0cm 0cm, width=0.45\textwidth]{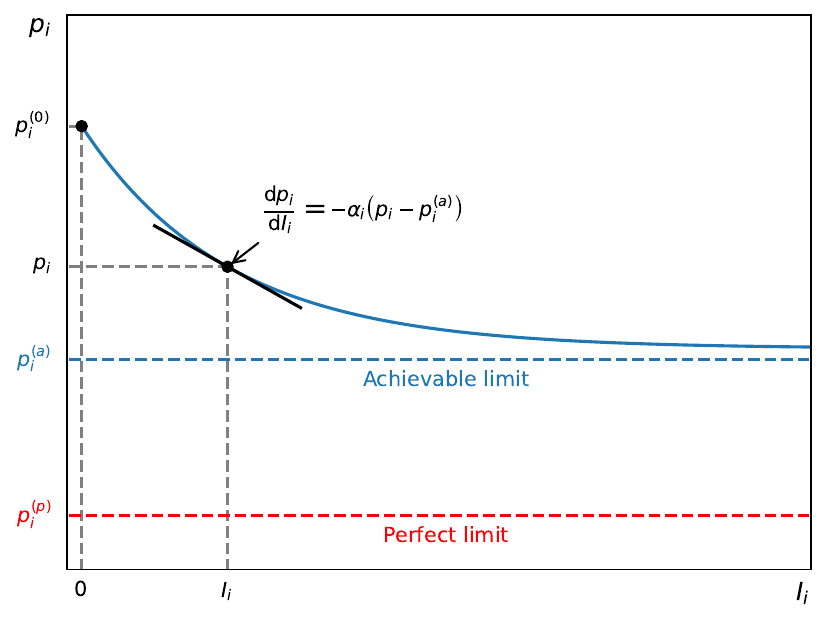}
    \caption{Representation of $p_{i}$ as a function of $I_{i}$. The slope of this function is customised by parameters $\alpha_{i}$ and $p_i^{(a)}$. The parameter $p_i^{(a)}$ is customised by parameter $\beta_{i}$ as shown in Eq.~(\ref{E:pia}).}
    \label{F:decay_function}
\end{figure}

The term $\frac{\mathrm{d}p_i}{\mathrm{d}I_i}$ is difficult to calculate, especially that $p_i$ as a function of $I_i$ depends on technology type and maturity as well as technical skills and business operation of storage developers. However, $\frac{\mathrm{d}p_i}{\mathrm{d}I_i}$ which represents the slope is expected to decrease as $I_{i}$ increases (see Fig.~\ref{F:decay_function}). Such reduction is expected as the potential of improvement is larger when $p_i$ is distant from the achievable limit. The \emph{achievable limit} is the best achievable value $p_{i}^{(a)}$ for parameter $p_{i}$, i.e., the highest achievable charge and discharge efficiencies and the lowest achievable costs of charge, discharge and energy storage capacities.

Therefore, a possible, and likely the simplest function that can represent $\frac{\mathrm{d}p_i}{\mathrm{d}I_i}$ is shown in Eq.~(\ref{E:dpi/dIi})

\begin{equation}
\label{E:dpi/dIi}
\frac{\mathrm{d}p_i}{\mathrm{d}I_i} = - \alpha_i \left (p_i - p_i^{(a)} \right )
\end{equation}

where $\alpha_i$ is a positive scaling parameter. When $p_{i}$ represents the charge or discharge efficiencies, the term $\left (p_i - p_i^{(a)} \right )$ is always negative and $p_{i}$ increases as $I_i$ increases. In contrast, when $p_{i}$ represents the charge, discharge, or energy capacity cost, the term $\left (p_i - p_i^{(a)} \right )$ is always positive and $p_{i}$ decreases as $I_i$ increases. To complete Eq.~(\ref{E:dpi/dIi}), $p_i^{(a)}$ is calculated as shown in Eq.~(\ref{E:pia})

\begin{equation}
\label{E:pia}
    p_{i}^{(a)} = \left( 1- \beta_i \right) p_{i}^{(0)} + \beta_i p_{i}^{(p)}
\end{equation}

where $\beta_i$ $\left(0\leq\beta_i\leq1\right)$ controls the $p_i^{(a)}$ location between $p_i^{(0)}$ and $p_i^{(p)}$ (see Fig.~\ref{F:decay_function}). The term $p_i^{(0)}$ represents the initial $p_i$ values, and the term $p_i^{(p)}$ is the perfect limit that corresponds to perfect charge and discharge efficiencies (no losses), and to no charge, discharge and energy capacity costs. When $p_i$ represents the charge and discharge efficiencies, $p_i^{(p)}$ is equal to $100\%$ except for the charge efficiency of heat pump technology; in this case the adopted $p_i^{(p)}$ value is $250\%$. When $p_i$ represents the capacity costs, $p_i^{(p)}$ is equal to zero.

The term $\frac{\mathrm{d}I_i}{\mathrm{d}I}$ is the decision variable to minimise the $LCOE$. For notation simplicity $\frac{\mathrm{d}I_i}{\mathrm{d}I}$ will be referred to as $\phi_i$. Consequently, Eq.~(\ref{E:dLCOE/dI}) can be rewritten as

\begin{equation}
\begin{aligned}
\label{E1:dLCOE/dI}
\frac{\mathrm{d}LCOE}{\mathrm{d}I} = &\sum_{i=1}^5 \frac{LCOE|_{p_i + \Delta p_i} - LCOE|_{p_i}}{\Delta p_i} \\
&\times \alpha_i \left[\left(1-\beta_i \right) p_{i}^{(0)} + \beta_i p_{i}^{(p)}- p_i\right ]\phi_i
\end{aligned} 
\end{equation}

The $LCOE$ minimisation based on Eq.~(\ref{E1:dLCOE/dI}) is difficult to solve; the difficulty is due to the interdependency of the gradient $(\frac{\partial g}{\partial p_i})$, $p_i$, and $\phi_i$. To overcome this difficulty, Eq.~(\ref{E1:dLCOE/dI}) is discretised and the minimisation is conducted for a series of investments $(\Delta I^{(n)})$. Hence, Eq.~(\ref{E1:dLCOE/dI}) can be rewritten as shown in 
Eq.~(\ref{E2:dLCOE/dI})

\begin{equation}
\begin{aligned}
\mathrm{\Delta}LCOE^{(n)} = & {\mathrm{\Delta}I^{(n)}}\sum_{i=1}^5 \frac{LCOE|_{p_i^{(n)} + \Delta p_i} - LCOE|_{p_i^{(n)}}}{\Delta p_i} \alpha_i\\
& \times \left[\left(1-\beta_i \right) p_{i}^{(0)} + \beta_i p_{i}^{(p)}- p_i^{(n)}\right ]\phi_i^{(n)}    
\end{aligned}
\label{E2:dLCOE/dI}
\end{equation}

where the superscript $(n)$ denotes the optimisation step. For every $n$ in the series of optimisation, $\phi_i^{(n)}$ is optimised and $\Delta I_i^{*(n)}$ values are calculated as shown in Eq.~(\ref{E:optimal values})

\begin{equation}
\label{E:optimal values}
\Delta I_i^{*(n)} = \phi_i^{*(n)}\Delta I^{(n)}    
\end{equation}

where $\Delta I_i^{*(n)}$ and $\phi_i^{*(n)}$ are the optimal investment and fraction of the optimal investment allocated to improving $p_i$ respectively. Based on $\Delta I_i^{*(n)}$, the integration of Eq.~(\ref{E:dpi/dIi}), and that the initial $p_{i}$ value is $p_{i}^{(0)}$, $p_i$ is updated as shown in Eq.~(\ref{E:pifunction})

\begin{equation}
\label{E:pifunction}
 p_{i}^{(n+1)} = p_i^{(a)} + \left(p_{i}^{(0)}-p_i^{(a)} \right)e^{-\alpha_i I_i^{*(n)}}
\end{equation}

where $I_i^*$ is the sum of previous investments allocated to improving $p_i$ as shown in Eq.~(\ref{E:I_i^*})

\begin{equation}
\label{E:I_i^*}
    I_i^{*(n)} = \sum_{n=0}^m \Delta I_i^{*(n)}
\end{equation}

The updated $p_i^{(n+1)}$ values from Eq.~(\ref{E:pifunction}) allows the iterative optimal allocation of investments to be repeated. In this iterative process, only positive investments are allowed, hence, $\phi_i$ should satisfy the constraint in Eq.~(\ref{E:alpha_constraints}). 
\begin{equation}
\label{E:alpha_constraints}
\phi_i^{(n)} \geq 0  ~~~~~~~~\forall i, \forall n
\end{equation}

\section{\label{S:Optimal improvements of emerging energy storage technologies}Optimal improvements of emerging energy storage technologies}
The parameters optimisation of storage technologies depends on $LCOE$ minimisation which is affected by various inputs that characterise the technical and financial characteristics of the hybrid renewable energy system. Inputs considered in this research to demonstrate the application of the investment-based optimisation method are presented in Section~\ref{S:Inputs}

\subsection{\label{S:Inputs}Inputs}

Four energy storage technologies are adopted in this research, namely, TES, pumped thermal energy storage (PTES), molten salt energy storage (MSES), or adiabatic compressed air energy storage (aCAES). In a TES system, electric heaters convert electrical energy to thermal energy that is stored in rocks or firebricks. When electricity is needed, the stored energy is extracted through a Brayton cycle that converts thermal energy to electricity~\cite{BENATO201829}. Unlike a TES system, a PTES system utilises electrical energy to pump thermal energy by a heat pump cycle from a cold tank to a hot tank. Both tanks store thermal energy in solid material such as volcanic rocks. The stored thermal energy is extracted through a Brayton cycle when electricity is needed~\cite{olympios2021progress, gautam2022review}.

An MSES system utilises electric heaters to converts electrical energy to thermal energy that is stored in tanks of molten salt. Unlike TES and PTES sytems, an MSES system produces electricity by discharging the stored thermal energy through a Rankine cycle~\cite{gautam2022review}. In an aCAES system, electrical energy is utilised to compress air to high-pressure during the charging the process. The compressed air is stored in either underground caverns or pressurised tanks. Thermal energy released during the compression process is also recovered and stored. During the discharge process, compressed air is released and reheated by the recovered thermal energy before entering the expander to generate electricity~\cite{GUO2019262}. More details about these technologies can be found in~\cite{BENATO201829}--\cite{GUO2019262}.

\begin{figure}
    \includegraphics[trim={0.0cm 0.0cm 0.0cm 0.0cm}, clip, width=0.5\textwidth]{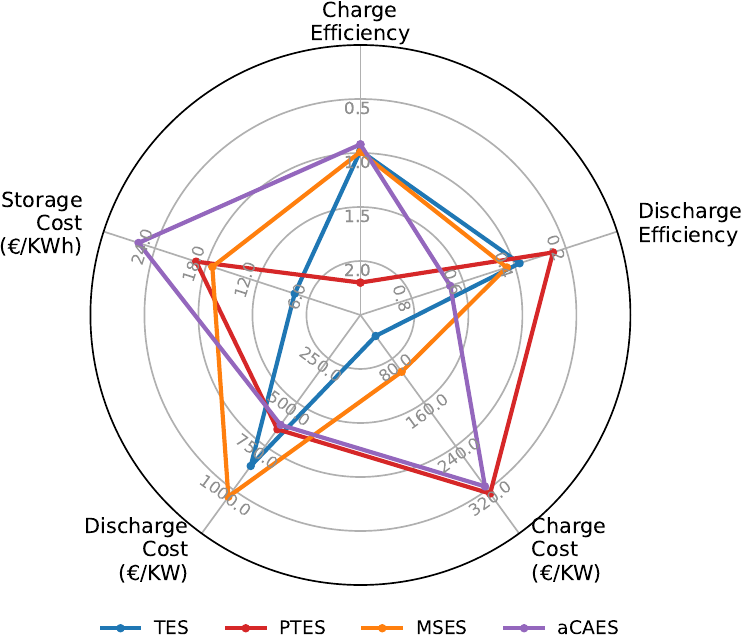}
    \caption{Parameters comparison of storage technologies.}
    \label{F:current_storages}
\end{figure}

Although other energy storage technologies could have also been included, the different characteristics of the adopted technologies are sufficient to demonstrate the application and usefulness of the optimisation method. The differences are highlighted in both the parameter values in Table~\ref{T:StorageX parameters} and in Fig.~\ref{F:current_storages} that shows the parameters considered for optimisation.

\begin{table}
\caption{\label{T:StorageX parameters}StorageX parameters \cite{GridScaleTechnologyExplained, gautam2022review, PRXEnergy.2.023006}.}
\begin{threeparttable}
\begin{ruledtabular}
\begin{tabular}{l c c c c c}
    Description & TES & PTES &  MSES & aCAES & Units\\
    \hline
    \textbf{Discharge} & \multirow{2}{*}{38\tnote{*}} & \multirow{2}{*}{25\tnote{*}} & \multirow{2}{*}{43} & \multirow{2}{*}{65} & \multirow{2}{*}{\%}\\
    \textbf{efficiency ($\eta_d$)} & & & & & \\
    \\[-0.75em]
    \textbf{Discharge} & \multirow{2}{*}{864} & \multirow{2}{*}{654} & \multirow{2}{*}{1040} & \multirow{2}{*}{628.8} & \multirow{2}{*}{$10^3$ \texteuro/MW}\\
    \textbf{capacity cost ($c_d$)} & & & & & \\
    \\[-0.75em]
    Discharge capacity & \multirow{2}{*}{17.28} & \multirow{2}{*}{13.08} & \multirow{2}{*}{20.80} & \multirow{2}{*}{12.58} & \multirow{2}{*}{$10^3$ \texteuro/MW}\\
    fixed cost ($f_d$) & & & & & \\
    \\[-0.75em]
    Discharge variable & \multirow{2}{*}{0} & \multirow{2}{*}{0} & \multirow{2}{*}{0} & \multirow{2}{*}{0} & \multirow{2}{*}{\texteuro/MWh}\\
    cost ($o_d$) & & & & & \\
    \\[-0.3em]
    \textbf{Charge} & \multirow{2}{*}{98} & \multirow{2}{*}{220} & \multirow{2}{*}{99} & \multirow{2}{*}{92} & \multirow{2}{*}{\%}\\
    \textbf{efficiency ($\eta_c$)} & & & & & \\
    \\[-0.75em]
    \textbf{Charge capacity} & \multirow{2}{*}{38.4} & \multirow{2}{*}{327} & \multirow{2}{*}{104} & \multirow{2}{*}{313.9} & \multirow{2}{*}{$10^3$ \texteuro/MW}\\
    \textbf{cost ($c_c$)} & & & & & \\
    \\[-0.75em]
    Charge capacity & \multirow{2}{*}{768} & \multirow{2}{*}{6540} & \multirow{2}{*}{2074} & \multirow{2}{*}{6278.4} & \multirow{2}{*}{\texteuro/MW}\\
    fixed cost ($f_c$) & & & & & \\
    \\[-0.75em]
    Charge variable & \multirow{2}{*}{0} & \multirow{2}{*}{0} & \multirow{2}{*}{0} & \multirow{2}{*}{0} & \multirow{2}{*}{\texteuro/MWh}\\
    cost ($o_c$) & & & & & \\
    \\[-0.3em]
    Energy loss ($L_s$) & 1 & 1 & 1 & 1 & \%\\
    \\[-0.75em]
    \textbf{Energy capacity} & \multirow{2}{*}{7.68} & \multirow{2}{*}{19.2} & \multirow{2}{*}{17.3} & \multirow{2}{*}{25.9} & \multirow{2}{*}{$10^3$ \texteuro/MWh}\\
    \textbf{cost ($c_s$)} & & & & & \\
    \\[-0.75em]
    Energy capacity & \multirow{2}{*}{153.6} & \multirow{2}{*}{384} & \multirow{2}{*}{345.6} & \multirow{2}{*}{518.4} & \multirow{2}{*}{\texteuro/MWh}\\
    fixed cost ($f_s$) & & & & & \\
    \\[-0.75em]
    Energy variable & \multirow{2}{*}{0} & \multirow{2}{*}{0} & \multirow{2}{*}{0} & \multirow{2}{*}{0} & \multirow{2}{*}{\texteuro/MWh}\\
    cost ($o_s$) & & & & & \\
    \\[-0.3em]
    Minimum state of & \multirow{2}{*}{20} & \multirow{2}{*}{20} & \multirow{2}{*}{20} & \multirow{2}{*}{20} & \multirow{2}{*}{\%}\\
    charge ($SOC_{4,min}$) & & & & &\\
    \\[-0.75em]
    Lifetime ($T_4$) & 40 & 40 & 40 & 30 & years\\
\end{tabular}
\begin{tablenotes}
    \item[*] The difference in efficiencies between TES and PTES systems that utilise a Brayton cycle for discharge, can be attributed to their distinct operating conditions 
\end{tablenotes}
\end{ruledtabular}
\end{threeparttable}
\end{table}

Fig.~\ref{F:current_storages} shows that the PTES has significantly higher charge efficiency, while aCAES has the highest discharge efficiency. Both PTES and aCAES have higher charge costs than the other technologies. The charge costs vary more than four-fold with TES having the lowest value. The MSES exhibits the highest discharge cost among the considered technologies. Storage costs vary almost four-fold, with aCAES and TES having the highest and lowest costs respectively. The comparison also reveals that aCAES has the worst charge efficiency and storage cost, while PTES has the worst discharge efficiency and charge cost. The MSES has the worst discharge cost among the considered storage technologies. This comparison suggests that a single investment strategy may not be suitable to improve the characteristics of all storage technologies.

In addition, each of the energy storage parameters in Fig.~\ref{F:current_storages} is likely to have its unique improvement characteristics defined by the corresponding $\alpha_i$ and $\beta_i$ values. These values can be customised for each parameter individually based on expected technological advancements, as well as possible operational and procurement improvements that could increase the potential of improving the efficiencies or reducing the capacity costs. The $\alpha_i$ and $\beta_i$ values are also technology specific, i.e., for the same energy storage parameter, the corresponding $\alpha_i$ and $\beta_i$ values are likely to be different for different energy storage technologies. The selection of appropriate $\alpha_i$ and $\beta_i$ values requires techno-economical expertise about energy storage technologies. However, for simplicity of demonstrating the application of the optimisation method, $\alpha_i$ and $\beta_i$ values are assumed to be equal as shown in Eq.~(\ref{E:constant_alpha_beta})

\begin{subequations}
\label{E:constant_alpha_beta}
\begin{align}
\alpha_i & = \alpha = 0.5 & \forall i\\
\beta_i & = \beta = 0.2 &\forall i
\end{align}
\end{subequations}

\begin{figure}
    \includegraphics[width=0.45\textwidth]{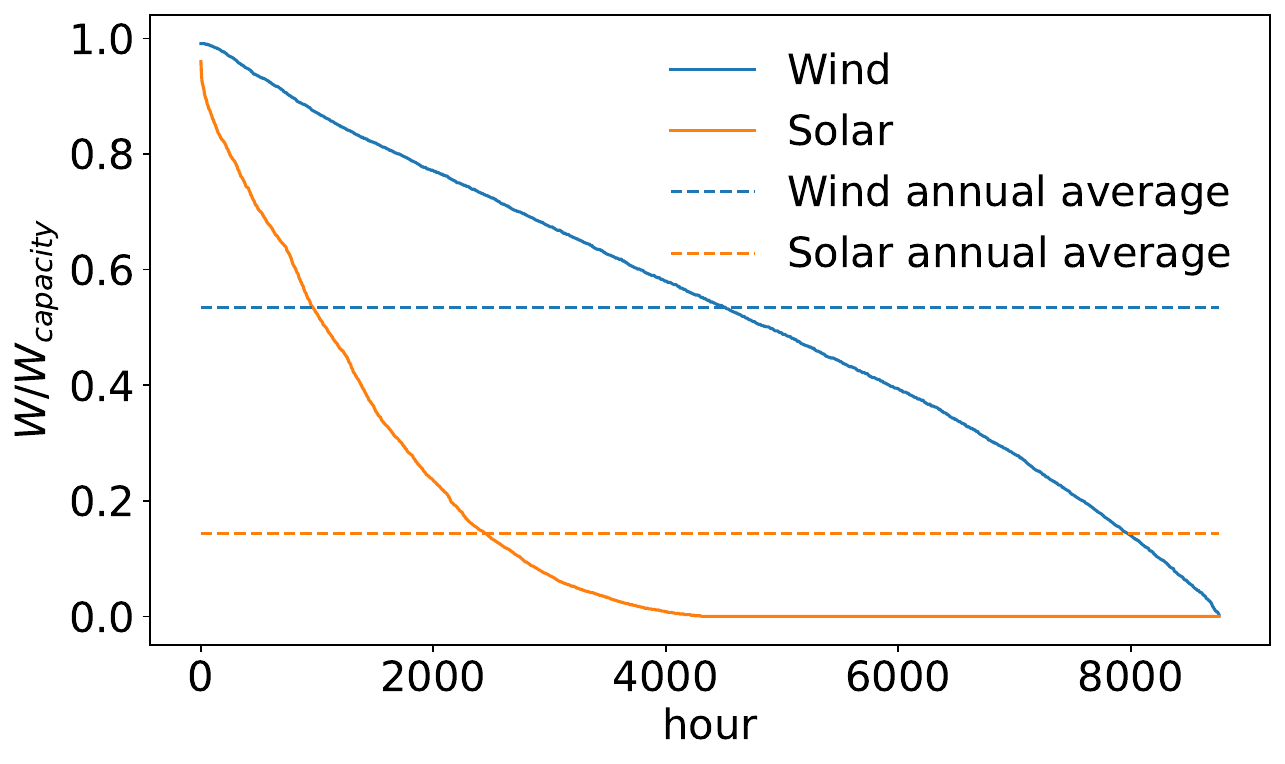}
    \caption{Solar and wind duration curves. The hourly capacity factors are shown in Fig.~\ref{F:SWCF}.}
    \label{F:SWDC}
\end{figure}

The optimisation results are also affected by the available solar and wind resources which are characterised by their respective capacity factors. The adopted capacity factors for solar and wind are shown in Appendix~\ref{A:Solar and wind capacity factors} in Fig.~\ref{F:SCF} and Fig.~\ref{F:WCF} respectively. The duration curves (Fig.~\ref{F:SWDC}) reveals that the potential of electricity production from wind systems is higher than the potential of electricity production from solar systems. The higher wind potential indicates that the wind capacity is likely to be higher than the solar capacity in an optimised hybrid system. For simplicity and for the purpose of focusing on the use of investment-based optimisation, the analysis does not consider solar and wind capacity factors that correspond to several geographical locations. Nevertheless, the results for additional years and for two extreme cases, without solar and without wind, are included in Appendix~\ref{B:Additional results}.

Battery and power converter are characterised by the parameters in Tables~\ref{T:Battery parameters}, \ref{T:Power converter parameters}, whereas the grid connection is characterised by a $100~MW$ capacity for electricity export and import and 0.85 capacity factor. The discount rate utilised in Eq.~(\ref{E:AC}) is set to 7\%. With these inputs specified, the optimisation of the storage parameters of the different technologies can be conducted. The optimisation results obtained based on these inputs are presented in Section~\ref{S:Results}. The results sensitivity to some of the input values are also considered; these results are presented in Appendix~\ref{B:Additional results} 

\begin{table}
\caption{\label{T:Battery parameters}Battery parameters~\cite{DanishEnergyAgency_storage}.}
\begin{ruledtabular}
\begin{tabular}{lcc}
 Description & Values & Units\\
    \hline
    Charge efficiency $(\eta_{3,c})$ & 98.5 & \%\\
    Disharge efficiency $(\eta_{3,d})$ & 97.5 & \%\\
    Daily energy loss $(L_{3})$ & 0.1 & \% \\
    \\[-0.75em]
    Storage to charge capacity & \multirow{2}{*}{2.0} & \multirow{2}{*}{MWh/MW}\\
    ratio $(E_3/G_{3,c})$ & & \\
    Discharge to charge capacity & \multirow{2}{*}{6.0} & \multirow{2}{*}{MW/MW}\\
    ratio $(G_{3,d}/G_{3,c})$ & & \\
    \\[-0.75em]
    Capacity cost $(c_{3})$ & 142 & $10^3\times$\texteuro/MWh\\
    Annual cost per discharge & \multirow{2}{*}{540} & \multirow{2}{*}{\texteuro/MW}\\
    capacity $(f_{3})$ & & \\
    Operation cost $(o_{3})$ & 1.8 & \texteuro/MWh\\
    \\[-0.75em]
    Minimum state of & \multirow{2}{*}{20} & \multirow{2}{*}{\%}\\
    charge $(SOC_{3,min})$\\
    Lifetime $(T_{3})$ & 25 & years\\
\end{tabular}
\end{ruledtabular}
\end{table}

\begin{table}
\caption{\label{T:Power converter parameters}Power converter parameters \cite{DanishEnergyAgency_inverter}.}
\begin{threeparttable}
\begin{ruledtabular}
\begin{tabular}{l c c}
    Description & Values & Units\\
    \hline
    Inverter efficiency ($\eta_{i}$) & 90\tnote{*} & \%\\
    Rectifier efficiency ($\eta_{r}$) & 90\tnote{*} & \%\\
    \\[-0.75em]
    Inverter capacity cost ($c_{i}$) & 20 & $10^3\times$\texteuro/MW\\
    Rectifier capacity cost ($c_{r}$) & 20 & $10^3\times$\texteuro/MW\\
    \\[-0.75em]
    Converter lifetime ($T_{c}$) & 15 & years\\
\end{tabular}
\begin{tablenotes}
    \item[*] Derated to account for part load operation
\end{tablenotes}
\end{ruledtabular}
\end{threeparttable}
\end{table}

\subsection{\label{S:Results}Results}
\begin{figure*}
  \begin{subfigure}[t]{0.49\textwidth}
    \includegraphics[width=\textwidth]{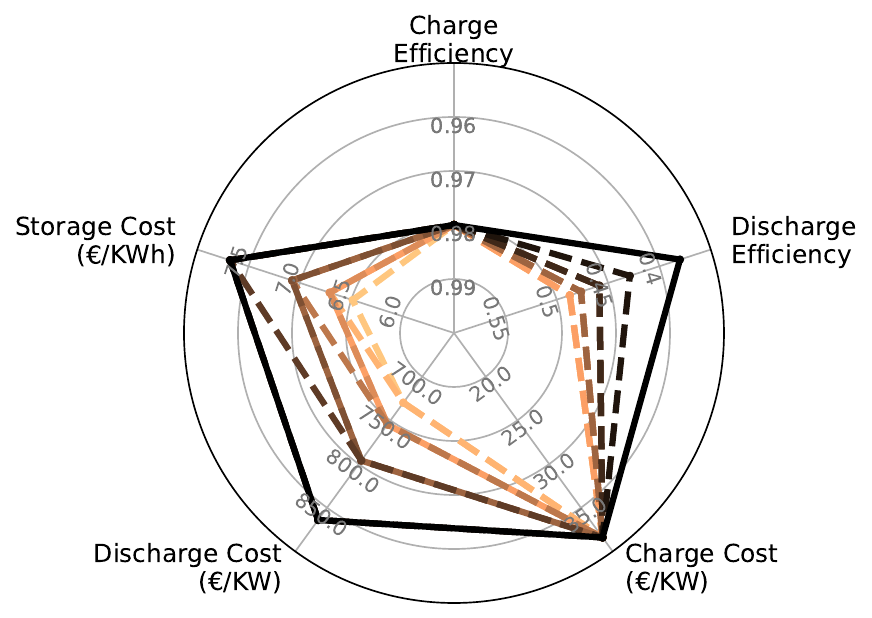}
    \caption{TES}
    \label{F:TES}
  \end{subfigure}
  \hfill
  \begin{subfigure}[t]{0.49\textwidth}
    \includegraphics[width=\textwidth]{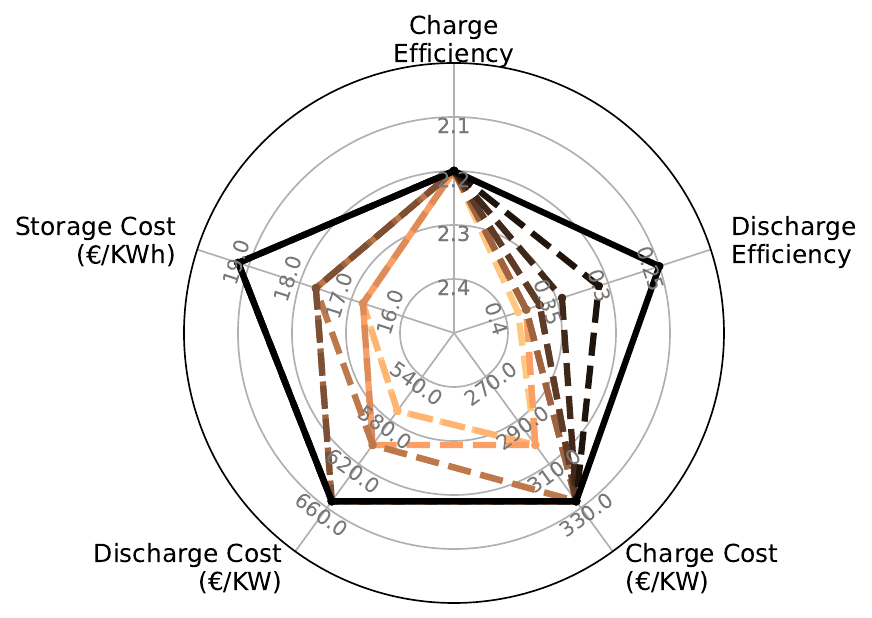}
    \caption{PTES}
    \label{F:PTES}
  \end{subfigure}

  \vspace{0.5cm}

  \begin{subfigure}[t]{0.49\textwidth}
    \includegraphics[width=\textwidth]{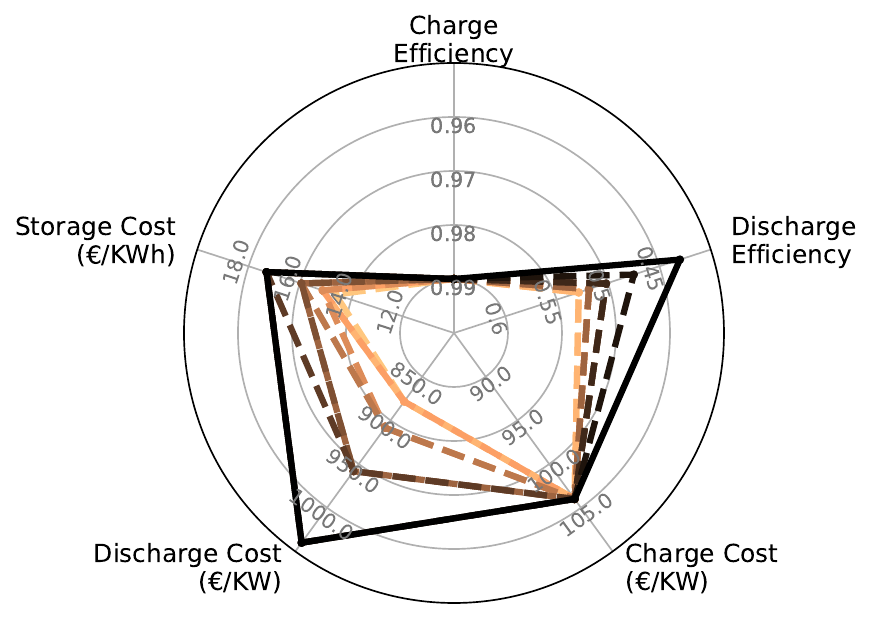}
    \caption{MSES}
    \label{F:MSES}
  \end{subfigure}
  \hfill
  \begin{subfigure}[t]{0.49\textwidth}
    \includegraphics[width=\textwidth]{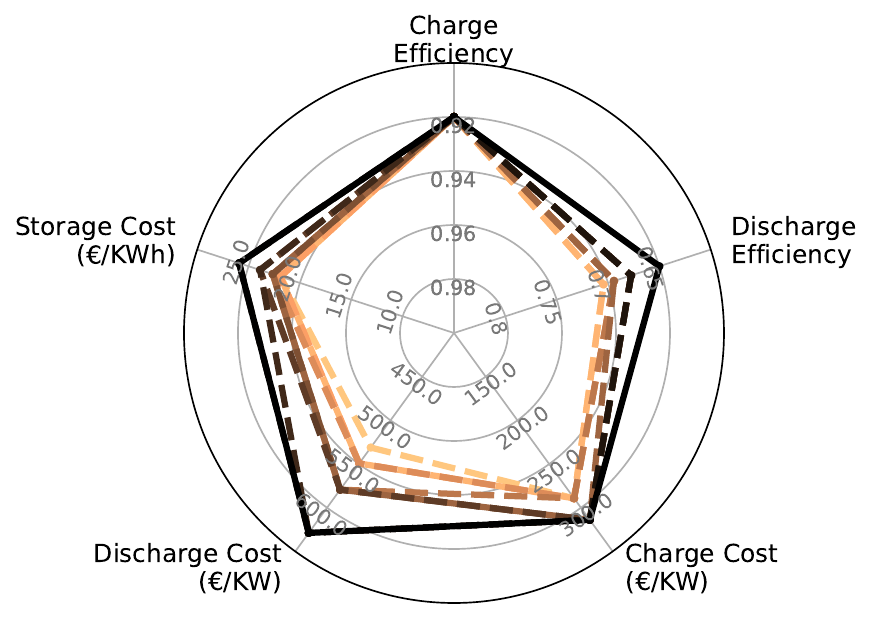}
    \caption{aCAES}
    \label{F:aCAES}
  \end{subfigure}

  \begin{subfigure}[t]{0.7\textwidth}
    \includegraphics[width=1.0\textwidth]{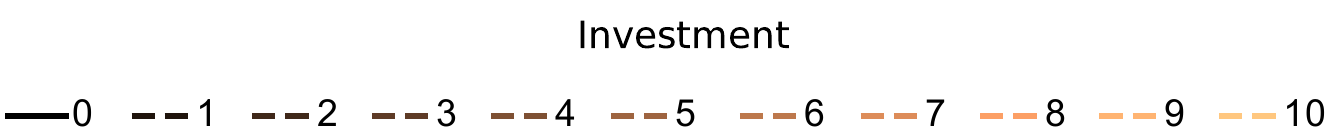}
  \end{subfigure}

  \caption{Optimal development path of storage technologies\hspace{\textwidth}
  year = 2019, grid connection capacity factor $CF_{G_7} = 0.85$,\hspace{\textwidth}
  slope parameter $\alpha=0.5$, improvement potential parameter $\beta=0.2$.}
  \label{F:Y2019-GCF0.85-alpha0.5-beta0.2}
\end{figure*}

The optimisation of storage technologies parameters is presented in Fig.~\ref{F:Y2019-GCF0.85-alpha0.5-beta0.2}. All the plots are interpreted similarly, with the most outer contour representing the initial efficiencies and cost parameters. Following the investments order, the plots show the corresponding parameters improvements. The results should be interpreted carefully as they are based on limited inputs that aim to demonstrate the usefulness of the investment-based optimisation method.

For TES technology (Fig.~\ref{F:TES}), the discharge efficiency is improved for the first two investments, followed by the discharge cost and then the storage cost. The subsequent investments alternate between these three parameters as they gradually become equally significant for the $LCOE$ minimisation. The charge efficiency and charge cost remain unchanged.

For PTES technology (Fig.~\ref{F:PTES}), the discharge efficiency is improved for the first few investments, followed by the storage cost and then the discharge cost. This order of improvements is slightly different compared to that of the TES. The subsequent investments alternate between these three parameters, however, the charge cost also improves towards the end of the investment series. The charge efficiency is the only parameter that remains unchanged.

The improvement patterns of MSES (Fig.~\ref{F:MSES}) and aCAES (Fig.~\ref{F:aCAES}) are similar to the improvements patterns of TES and PTES respectively. However, the aCAES improvements alternate among the discharge efficiency, storage cost, and discharge cost from the beginning of the investment series. This improvement pattern differs slightly from the PTES improvement pattern whereby the first few investments are dedicated to improving the discharge efficiency.

\begin{figure*}
  \begin{subfigure}[t]{0.49\textwidth}
    \includegraphics[trim={0.0cm 0.0cm 0.0cm 0.0cm}, clip, width=\textwidth]{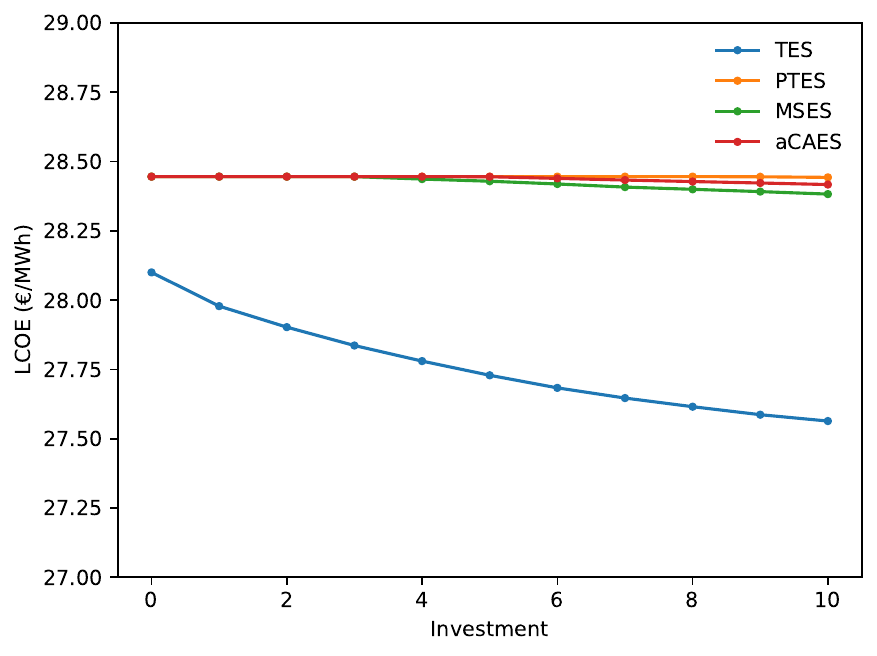}
    \caption{LCOE}
    \label{F:Effect of storage improvement on LCOE}
  \end{subfigure}
  \hfill
  \begin{subfigure}[t]{0.49\textwidth}
    \includegraphics[trim={0.0cm 0.0cm 1.5cm 0.0cm}, clip, width=\textwidth]{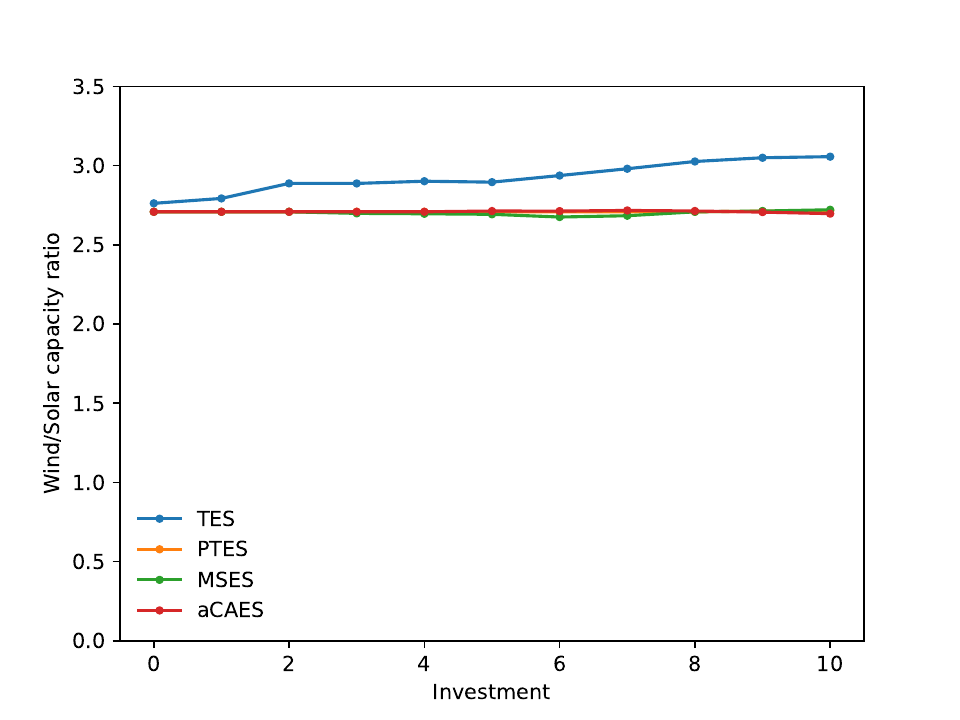}
    \caption{Wind/Solar capacity ratio}
    \label{F:Wind_over_solar_capacity_ratio}
  \end{subfigure}
  \vspace{0.5cm}
  \begin{subfigure}[t]{0.49\textwidth}
   \includegraphics[trim={0.0cm 0.0cm 0.0cm 0.0cm}, clip, width=\textwidth]{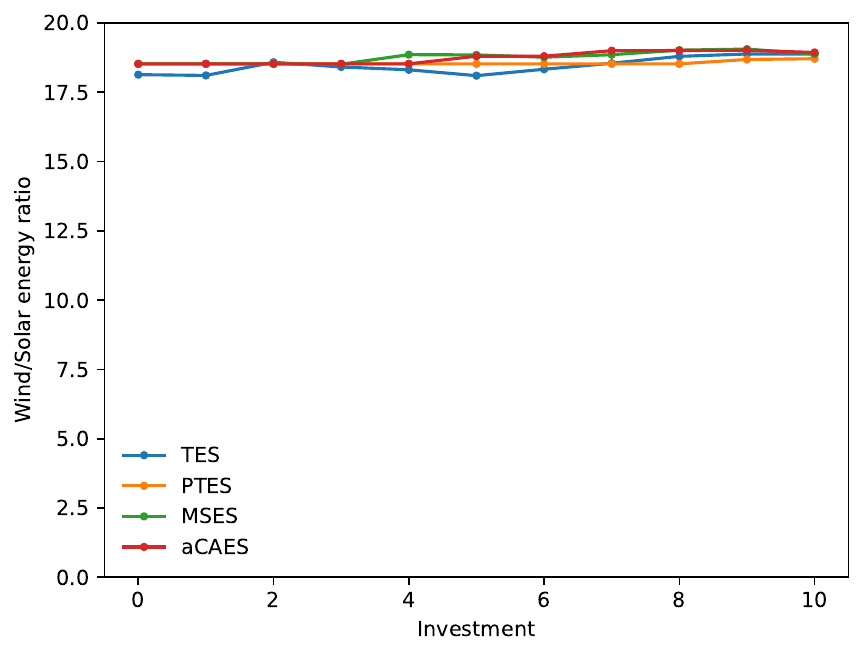}
    \caption{Wind/Solar energy ratio}
    \label{F:Wind_over_solar_energy_ratio}
  \end{subfigure}
  \hfill
  \begin{subfigure}[t]{0.49\textwidth}
    \includegraphics[trim={0.0cm 0.0cm 0.0cm 0.0cm}, clip, width=\textwidth]{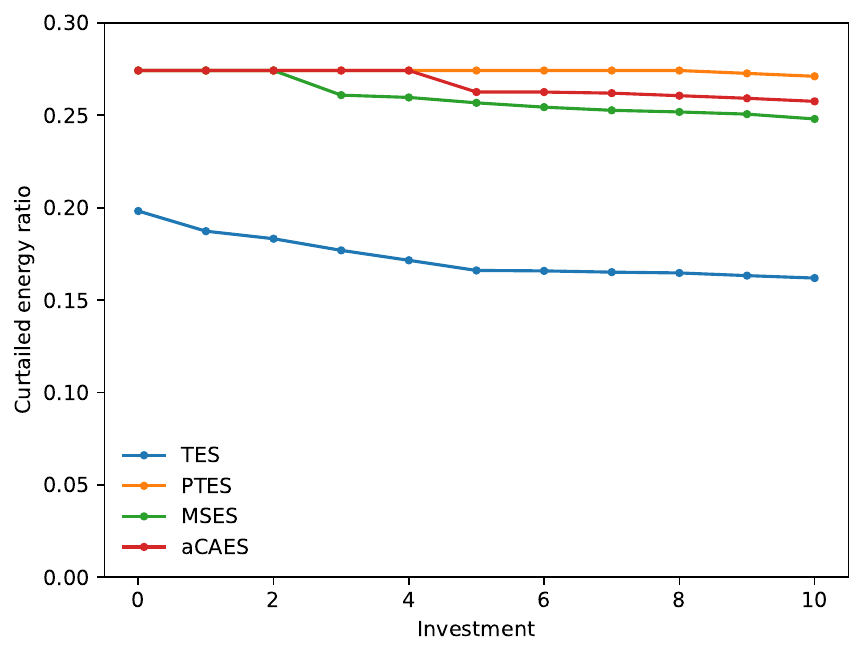}
    \caption{Curtailed energy ratio}
    \label{F:Curtailed energy ratio}
  \end{subfigure}
  \caption{Effect of storage improvements on (a) LCOE, (b) Wind/Solar capacity ratio, (c) Wind/Solar energy ratio, and (d) Curtailed energy ratio.}
  \label{F:OptimalPath}
\end{figure*}

The improvements of the storage technologies reduce the $LCOE$ as shown in Fig.~\ref{F:Effect of storage improvement on LCOE}. The investments for improving the parameters of TES, which is already part of the optimal system configuration, achieve a noticeable effect on reducing the $LCOE$. However, the first few investments for improving the parameters of PTES, MSES, and aCAES have no effect on reducing the $LCOE$. Despite the initial improvements, these technologies remain excluded from the optimal system configuration. As the characteristics of MSES and aCAES continue to improve a small $LCOE$ reduction can be observed. However, despite the improvements of PTES parameters through the series of investments, the $LCOE$ remains practically unchanged.

In addition, the improvements of the storage technologies seem to have a limited effect on the ratios of solar and wind power capacity, utilised energy, and curtailed energy as shown in Fig.~\ref{F:Wind_over_solar_capacity_ratio}-\ref{F:Curtailed energy ratio} respectively. The wind/solar capacity ratio remains practically constant as the characteristics of the storage technologies improve. The wind/solar energy ratio has an increasing trend as the TES improves. However, despite the upward trend, the change of the wind/solar energy ratio is relatively small. The wind/solar energy ratio for PTES, MSES, and aCAES technologies is practically unaffected by the improvements. The effect of improving the storage technologies is more apparent for the curtailed energy ratio as shown in Fig.~\ref{F:Curtailed energy ratio}. This ratio is the total curtailed energy from both solar and wind divided by their total energy production. Improving the storage technologies reduces the curtailed energy ratio by a few percentage points for TES, as well as for MSES and aCAES after significant improvements. The curtailed energy ratio is practically unaffected by the PTES improvements.

To analyse the sensitivity of the investment optimisation results additional scenarios are presented in Appendix~\ref{B:Additional results}. The additional scenarios consider different time series for solar and wind capacity factors, different capacity factors for the grid connection, and different $\alpha$ and $\beta$ values.

\section{\label{S:Discussion}Discussion}

The aim of this discussion is to identify the order of importance of storage parameters that energy storage developers should consider for improving their technologies. The discussion considers the results presented in both Section~\ref{S:Optimal improvements of emerging energy storage technologies} and Appendix~\ref{B:Additional results}. The interpretations of results presented in Appendix~\ref{B:Additional results} are similar to the results interpretations presented in section~\ref{S:Optimal improvements of emerging energy storage technologies}. The discussion also includes a comparison between the results obtained in this research and the results in \cite{sepulveda2021design, PRXEnergy.2.023006}.

The investment optimisation shows that the sequence of investments to improve the storage parameters may change as the solar and wind capacity factors change (Fig.~\ref{F:Y2019-GCF0.85-alpha0.5-beta0.2}, Fig.~\ref{F:Y2018-GCF0.85-alpha0.5-beta0.2}--Fig.~\ref{F:Y2019-GCF0.85-alpha0.5-beta0.2-NoWind}). The sequence of investments reveal that improving the discharge efficiency should be the first priority for energy storage developers. In contrast, improving the charge efficiency can be ignored. The second and third priorities for energy storage developers is to reduce either the discharge cost or the storage cost. The fourth priority is to reduce the charge cost. However, for renewable energy systems without wind power generation, reducing the charge cost becomes more important than reducing the discharge cost for all storage technologies except TES. This change of the order of importance can be explained by the significant increase of charge capacity when wind power generation is unavailable. When wind power generation is part of the renewable power system, the charge and discharge capacities are of the same order of magnitude. However, without wind power generation, a large energy storage capacity is needed, and a large charge capacity that is multiple times larger than the discharge capacity is also needed. The larger charge capacity increases the importance of reducing the charge cost. The TES exception is due to the initial charge cost being significantly small.

In addition, the capacity factor of the grid connection can affect the energy system configuration and the investments sequence (Fig.~\ref{F:Y2019-GCF0.85-alpha0.5-beta0.2}, Fig.~\ref{F:Y2019-GCF0.75-alpha0.5-beta0.2}--Fig.~\ref{F:Y2019-GCF0.95-alpha0.5-beta0.2}). For the lower capacity factor of the grid connection, the optimal system configuration includes neither the storageX technologies nor the battery; the TES becomes part of the optimal system only towards the end of the improvements sequence. The discharge efficiency remains the most important parameter to improve. The order of importance of other parameters depends on the storage technology. As the capacity factor of the grid connection increases, the improvement priority is the discharge efficiency followed by either the discharge cost or the storage cost. The least important parameters to improve are the charge cost and charge efficiency. The exception to this priority order is the aCAES storage cost being more important than the discharge efficiency. This exception at a high capacity factor of the grid connection is due to both the need for large energy storage capacity and to the initial discharge efficiency being relatively high. The high initial discharge efficiency reduces the benefits from further improvements, making the reduction of storage cost more important.

The overall results indicate that improving the discharge efficiency is the most significant for energy storage developers to reduce the LCOE. The second and third most important parameters are the discharge cost and the storage cost. Both parameters can be considered as equally important; the order of investments sequence for these two parameters depends on the storage technology and the operation conditions. The least important parameters are the charge cost and the charge efficiency. The initial charge efficiencies are already close to perfect efficiency, hence, improving the charge efficiency any further should be avoided by energy storage developers.

These results show some similarities as well as differences with the results in~\cite{sepulveda2021design, PRXEnergy.2.023006} (see Table~\ref{T:Storage parameter importance in descending order}). All the results indicate that the discharge efficiency and the storage cost are important parameters. However, the results obtained in this research indicate that the discharge efficiency is more important than the storage cost, whereas the results in~\cite{sepulveda2021design, PRXEnergy.2.023006} indicate that either the storage cost is the most important parameter or the two parameters are equally important.

\begin{table}[b]
\caption{\label{T:Storage parameter importance in descending order}Storage parameter significance in descending order}
\begin{ruledtabular}
\begin{tabular}{ccc }
System & Storage energy & $LCOE$\\
value~\cite{sepulveda2021design} & capacity deployment~\cite{PRXEnergy.2.023006} & (current results)\\
\colrule
    $c_{s}$ & \multirow{2}{*}{$c_{s}$, $\eta_{d}$} & $\eta_{d}$\\
    $\eta_{d}$ & & \multirow{2}{*}{$c_{d}$, $c_{s}$}\\
    $c_{d}$ & $c_{c}$\\
    $\eta_{c}$ & \multirow{2}{*}{$\eta_{c}$, $c_{d}$} & $c_{c}$\\
    $c_{c}$ & & $\eta_{c}$\\
\end{tabular}
\end{ruledtabular}
\end{table}

The discharge cost is one of the relatively important parameters according to the results obtained in this research and in~\cite{sepulveda2021design}. However, the results in~\cite{PRXEnergy.2.023006} show that the discharge cost is one of the least important parameters. The charge cost is one of the least important parameters which seems inline with the results in~\cite{sepulveda2021design} that show the charge cost as the least important parameter. The results in~\cite{PRXEnergy.2.023006} show that the charge cost is a relatively important parameter. The charge efficiency is the least important parameter as per the results in this research, and one of the least important parameters as per the results in~\cite{sepulveda2021design, PRXEnergy.2.023006}.

The comparison with results in~\cite{sepulveda2021design, PRXEnergy.2.023006} shows that in general, energy storage developers should initially attempt to improve the discharge efficiency and the storage cost. The remaining parameters can be ignored unless the energy storage developers have sufficient evidence that one of the remaining parameters have a priority for improvement. The comparison also reveals discrepancies of the precise order of importance of the storage parameters between the different studies. The discrepancies might be due to the energy system configuration, the operation conditions, or the methods utilised to identify the important parameters. The precise order of importance is critical for energy storage developers that undoubtedly prefer to be confident that their investments are {\it optimally} improving a specific energy storage technology. The methods and consequently the results in~\cite{sepulveda2021design, PRXEnergy.2.023006} cannot provide the precise order of importance of the storage parameters for specific technologies. However, the method and results obtained from this research are technology specific and provide precise order of importance of the storage parameters.

Nevertheless, the results obtained in this research should be interpreted carefully as they are based on a simple case study. For instance, the system does not consider sector coupling that could influence the role and importance of energy storage in the system~\cite{GEABERMUDEZ2021116685}. The $\alpha_i$ and $\beta_i$ parameters that control the investment effect on improvements and the potential of improvements are assumed constant for all the energy storage parameters. Estimating different $\alpha_i$ and $\beta_i$ values accurately for each energy storage parameter requires techno-economic expertise in energy storage development. The estimation of $\alpha_i$ and $\beta_i$ values may be a challenging aspect of utilising the investment-based optimisation method, however, the possibility of specifying these values is also the important feature of this method that allows energy storage developers specify the $\alpha_i$ and $\beta_i$ values according to both their energy storage technology and their skills in improving the energy storage parameters. It is also worth mentioning that the investment-based optimisation method is unrestricted to analysing energy storage technologies and can applied to different parameters and systems.

\section{\label{S:Summary and conclusions}Summary and conclusions}

This research presented a novel method to optimise the parameters of four energy storage technologies, namely, thermal energy, pumped thermal energy, molten salt, and adiabatic compressed air. These storage technologies were considered in a grid-connected hybrid renewable energy system that included solar and wind energy sources, and a lithium-ion battery. The energy system had to deliver a minimum annual amount of energy to the grid based on a specified capacity factor of the grid connection. The energy system was optimised with the objective of minimising the levelised cost of energy ($LCOE$).

The investment-based optimisation method split the relatively difficult optimisation of the energy storage parameters into two linear optimisation problems. In an iterative process, the first optimisation calculated the optimal $LCOE$ of the renewable energy system, while the second optimisation calculated the optimal distribution of investments to improve the energy storage parameters. In this iterative process, the optimisation method considered the improvement potential of the energy storage parameters. The investments sequence revealed the parameters hierarchy for minimising the $LCOE$. The adopted method provided a detailed order of parameters importance for different operation conditions for each of the energy storage technologies. The overall results indicated that the discharge and charge efficiencies are the most and least important parameters to minimise the $LCOE$ respectively. To some extent, the overall results shared some similarities and differences with the results from the existing literature. 

The results suggest that significant improvements of the emerging energy storage technologies are needed to outperform lithium-ion batteries. This research provides a method for developers to optimally improve energy storage technologies. The results are meant to demonstrate the capabilities of the investment-based method and to complement the results in the literature by considering a smaller renewable energy system. Future research may investigate if the discrepancy with the existing literature is due to the significant difference of the analysed energy systems. Such investigation can be achieved by applying the investment-based optimisation method to energy storage technologies in the systems described in the literature. Future research may also investigate the impact of different values for the parameters that control the investment optimisation. These parameters could possibly be linked to technology readiness levels; a high technology readiness level indicates a low potential for improvement. These potential future investigations could consider also consider other emerging energy storage technology that were not considered in this research.

Finally, large energy system models indicate that energy storage could be essential for the transition to a green economy. Therefore, a sustained effort of improving techno-economic characteristics of energy storage technologies is important to achieve the green transition. 

\begin{acknowledgments}

This research is funded by the GridScale project supported by the Danish Energy Technology Development and Demonstration Program under grant number 64020-2120. Dr Farah would like to thank Mr Ebbe Kyhl Gøtske for the interesting discussions during the development of this research.

\end{acknowledgments}

\appendix

\section{\label{A:Solar and wind capacity factors}Solar and wind capacity factors}

\begin{figure}
  \centering
  \begin{subfigure}[t]{0.45\textwidth}
    \includegraphics[width=\textwidth]{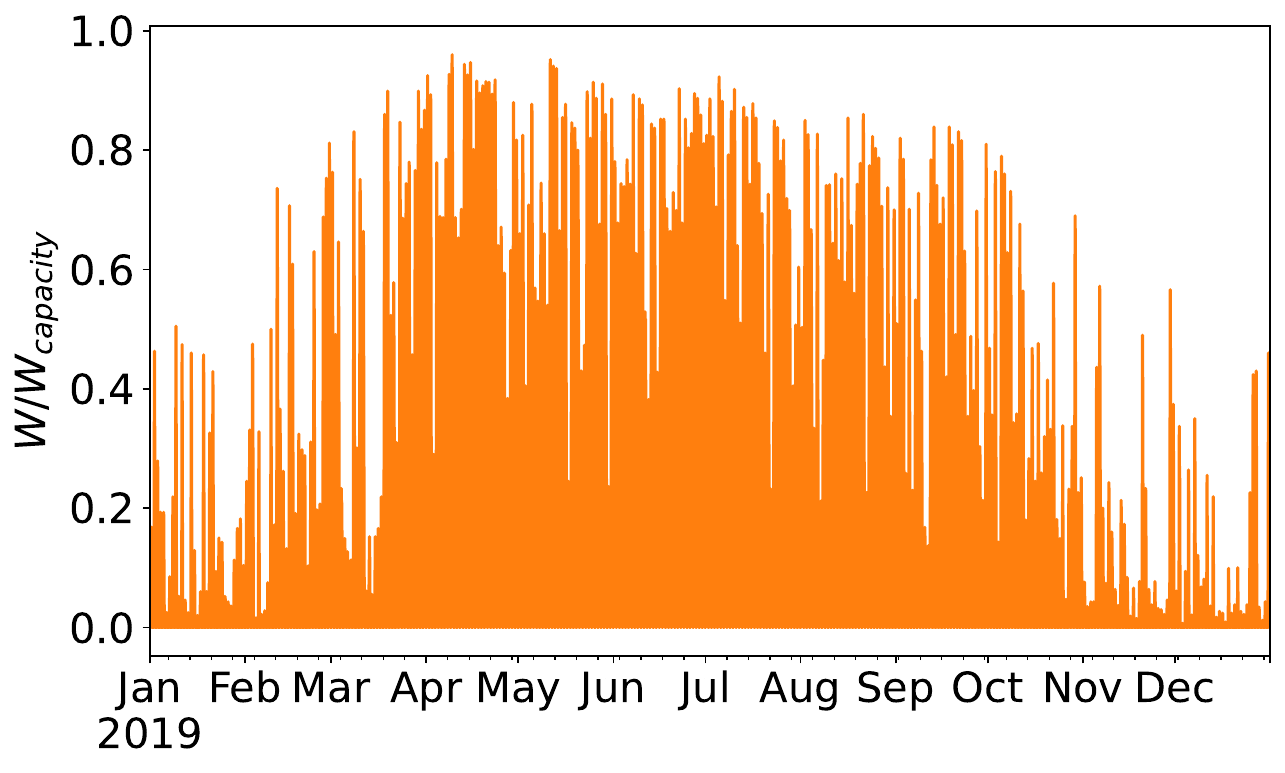}
    \caption{Solar capacity factor time series ($CF_{G_1}$). System loss (fraction): 0.0, tracking: None, tilt: 40\degree,  azimuth: 180\degree.}
    \label{F:SCF}
  \end{subfigure}
  \hfill
  \begin{subfigure}[t]{0.45\textwidth}
    \includegraphics[width=\textwidth]{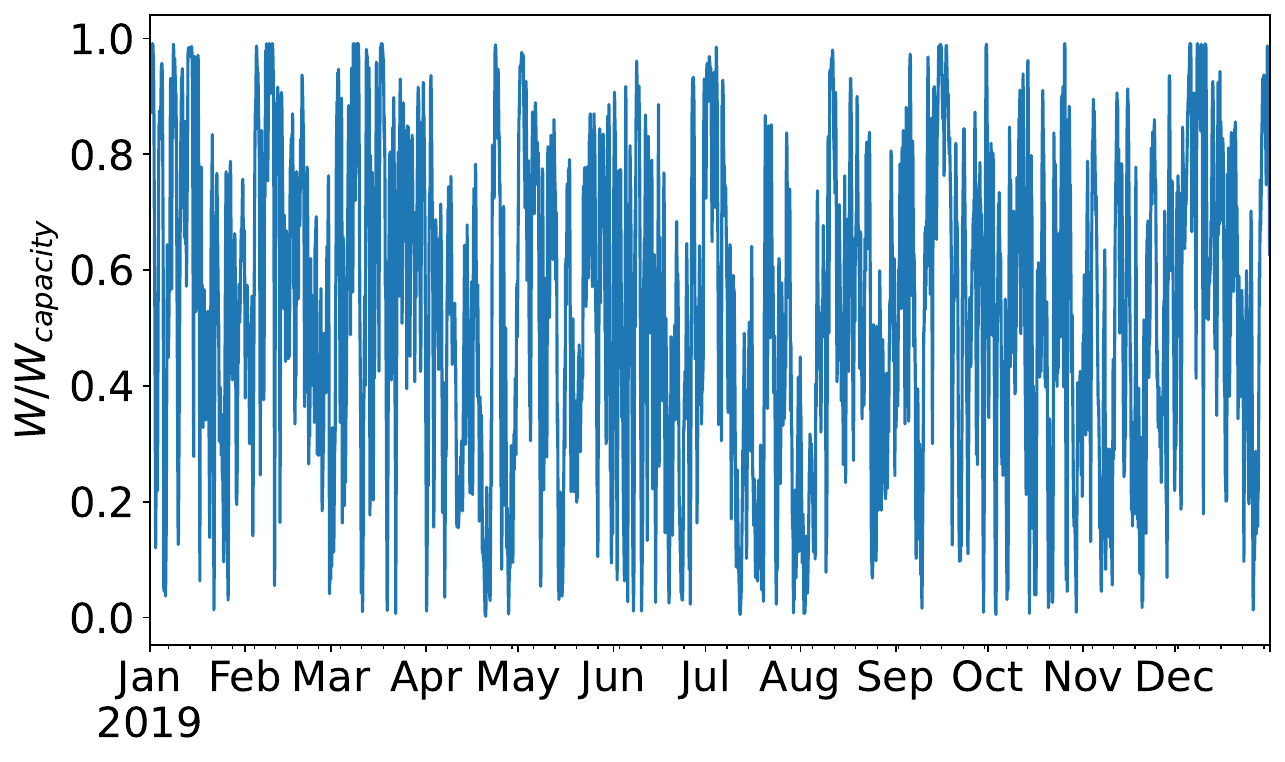}
    \caption{Wind capacity factor time series ($CF_{G_2}$). Hub height: 80.0~m, turbine model: Vestas V90 2000.}
    \label{F:WCF}
  \end{subfigure}
  \caption{Solar and wind capacity factors. Dataset: MERRA-2 (global), year: 2019, capacity: 1~kW, latitude: 56.6135\degree, longitude: 8.9328\degree.}
  \label{F:SWCF}
\end{figure}

Solar and wind capacity factors that correspond to the results in Section~\ref{S:Results} are shown in Fig.~\ref{F:SCF} and Fig.~\ref{F:WCF} respectively. Capacity factors for 2018 and 2020 that correspond to a part of the results in Appendix~\ref{B:Additional results} are not included.

\section{\label{B:Additional results}Additional results}
Results of additional scenarios are presented to assist in evaluating the sensitivity of storage parameters to solar and wind capacity factors, grid capacity factor, and the investment optimisation parameters $\alpha$ and $\beta$.

\begin{figure*}
  \centering
  \begin{subfigure}[t]{0.49\textwidth}
    \centering
    \includegraphics[width=\textwidth]{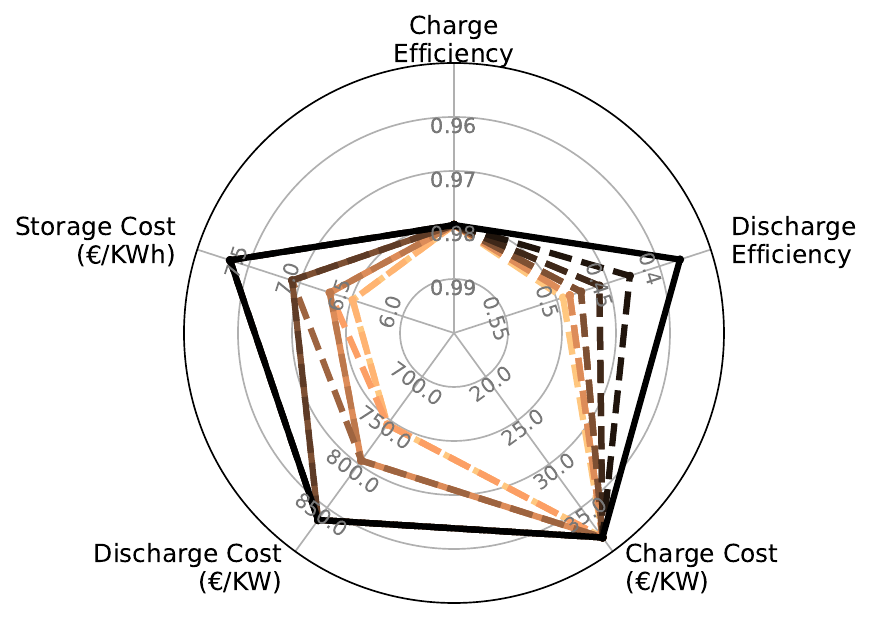}
    \caption{TES}
    \label{F:TES-Y2018-GCF0.85-alpha0.5-beta0.2}
  \end{subfigure}
  \hfill
  \begin{subfigure}[t]{0.49\textwidth}
    \centering
    \includegraphics[width=\textwidth]{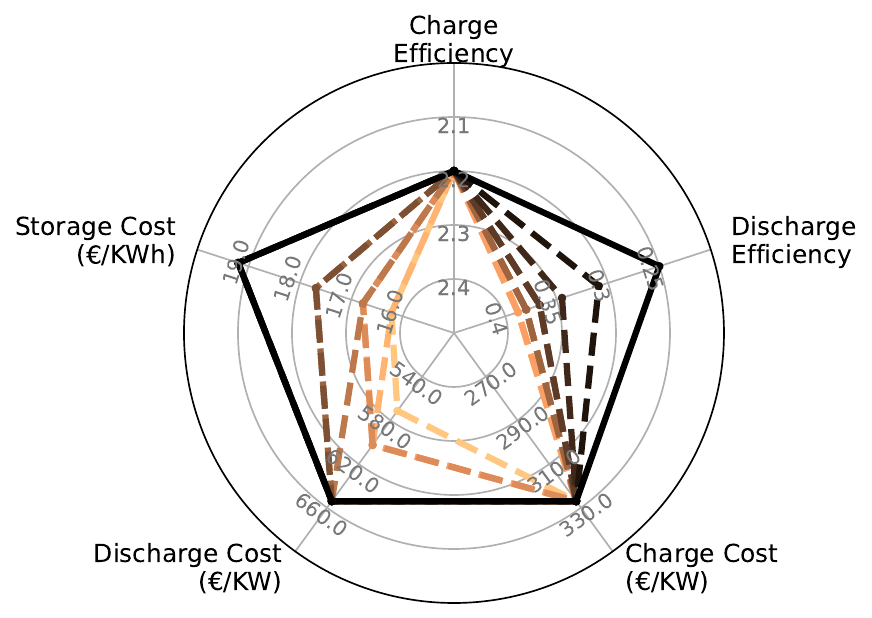}
    \caption{PTES}
    \label{F:PTES-Y2018-GCF0.85-alpha0.5-beta0.2}
  \end{subfigure}

  \vspace{0.5cm}

  \begin{subfigure}[t]{0.49\textwidth}
    \centering
    \includegraphics[width=\textwidth]{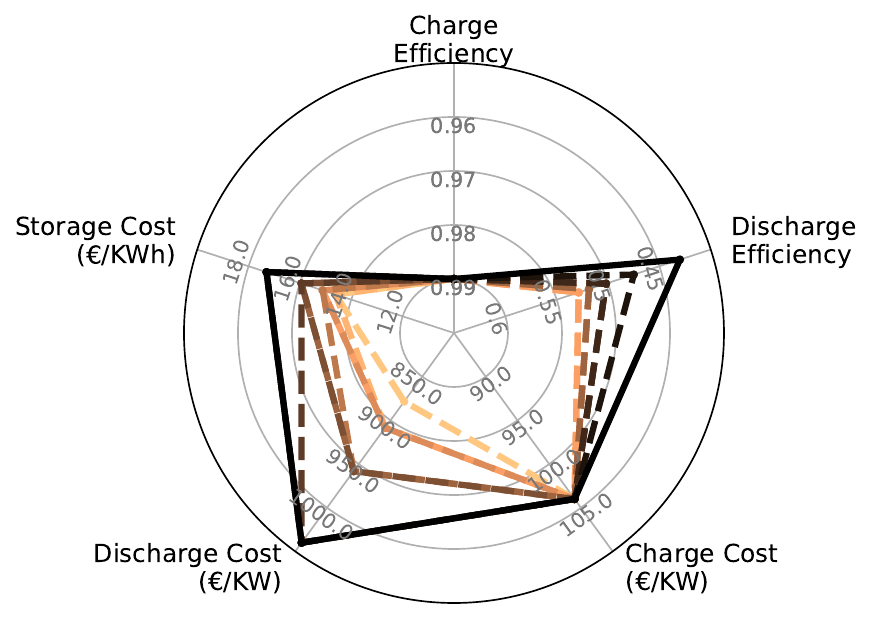}
    \caption{MSES}
    \label{F:MSES-Y2018-GCF0.85-alpha0.5-beta0.2}
  \end{subfigure}
  \hfill
  \begin{subfigure}[t]{0.49\textwidth}
    \centering
    \includegraphics[width=\textwidth]{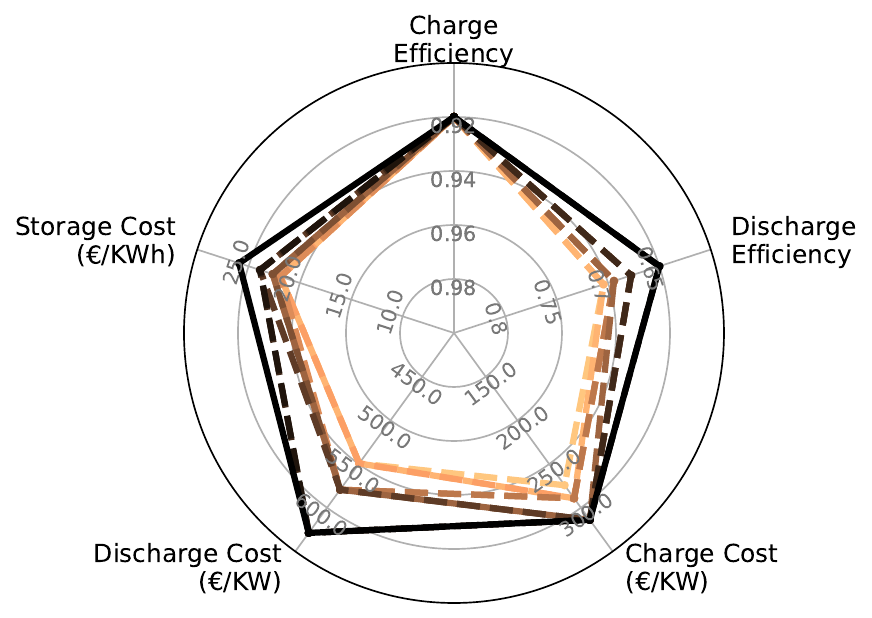}
    \caption{aCAES}
    \label{F:aCAES-Y2018-GCF0.85-alpha0.5-beta0.2}
  \end{subfigure}

  \begin{subfigure}[t]{0.7\textwidth}
    \centering
    \includegraphics[width=1.0\textwidth]{Figures/Denmark/Legend_2.pdf}
  \end{subfigure}

  \caption{Optimal development path of storage technologies \hspace{\textwidth}
  year = 2018, grid connection capacity factor $CF_{G_7}=0.85$, \hspace{\textwidth}
  slope parameter $\alpha=0.5$, improvement potential parameter $\beta=0.2$.}
  \label{F:Y2018-GCF0.85-alpha0.5-beta0.2}
\end{figure*}

\begin{figure*}
  \centering
  \begin{subfigure}[t]{0.49\textwidth}
    \centering
    \includegraphics[width=\textwidth]{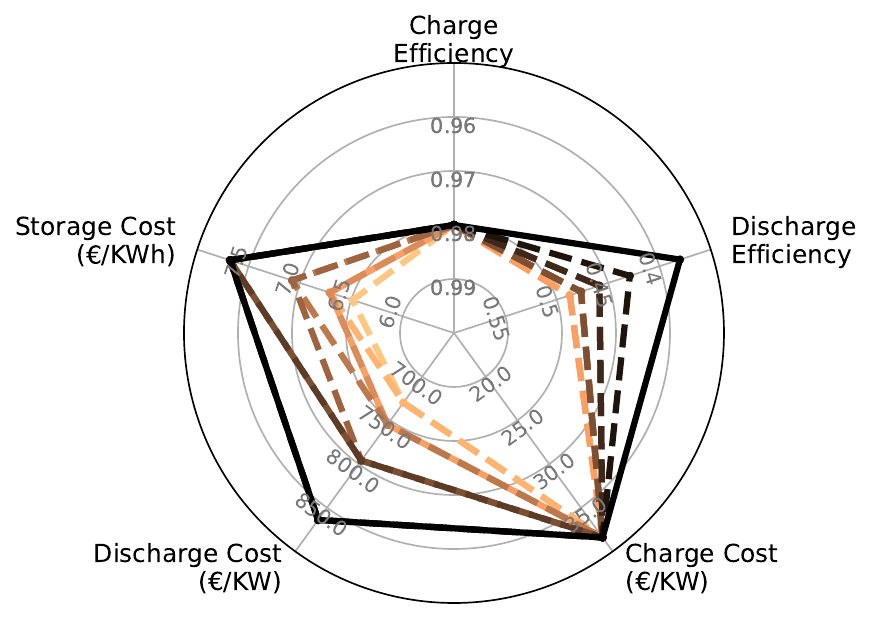}
    \caption{TES}
    \label{F:TES-Y2020-GCF0.85-alpha0.5-beta0.2}
  \end{subfigure}
  \hfill
  \begin{subfigure}[t]{0.49\textwidth}
    \centering
    \includegraphics[width=\textwidth]{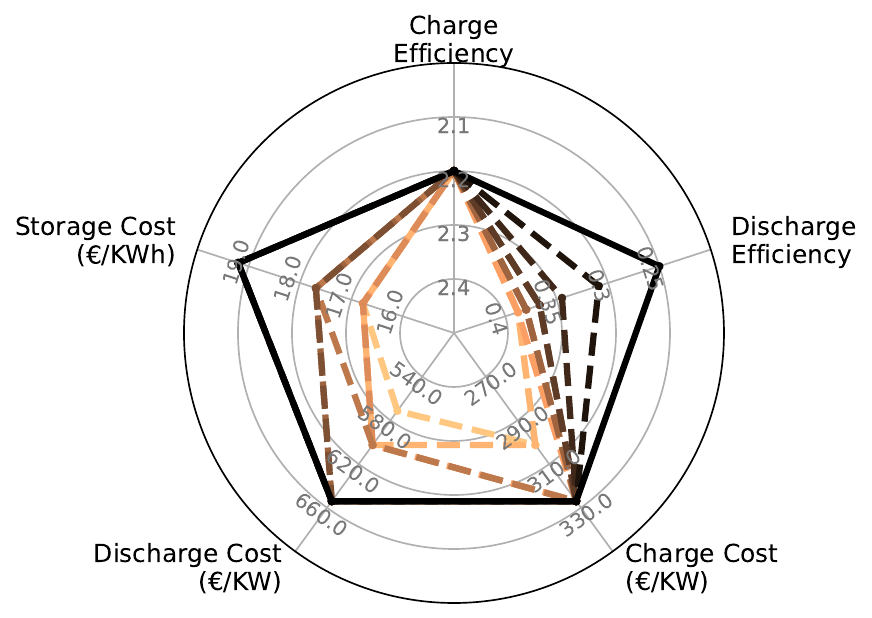}
    \caption{PTES}
    \label{F:PTES-Y2020-GCF0.85-alpha0.5-beta0.2}
  \end{subfigure}

  \vspace{0.5cm}

  \begin{subfigure}[t]{0.49\textwidth}
    \centering
    \includegraphics[width=\textwidth]{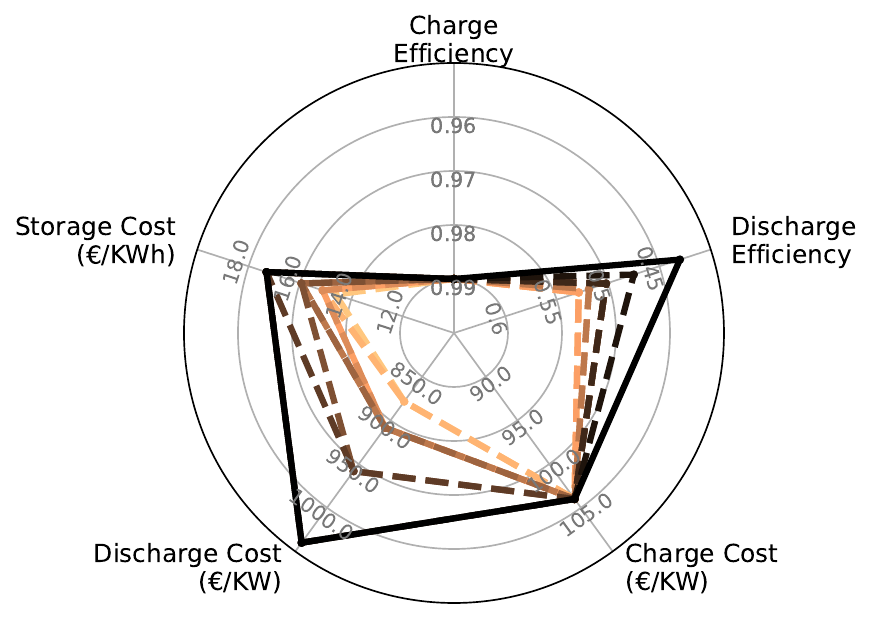}
    \caption{MSES}
    \label{F:MSES-Y2020-GCF0.85-alpha0.5-beta0.2}
  \end{subfigure}
  \hfill
  \begin{subfigure}[t]{0.49\textwidth}
    \centering
    \includegraphics[width=\textwidth]{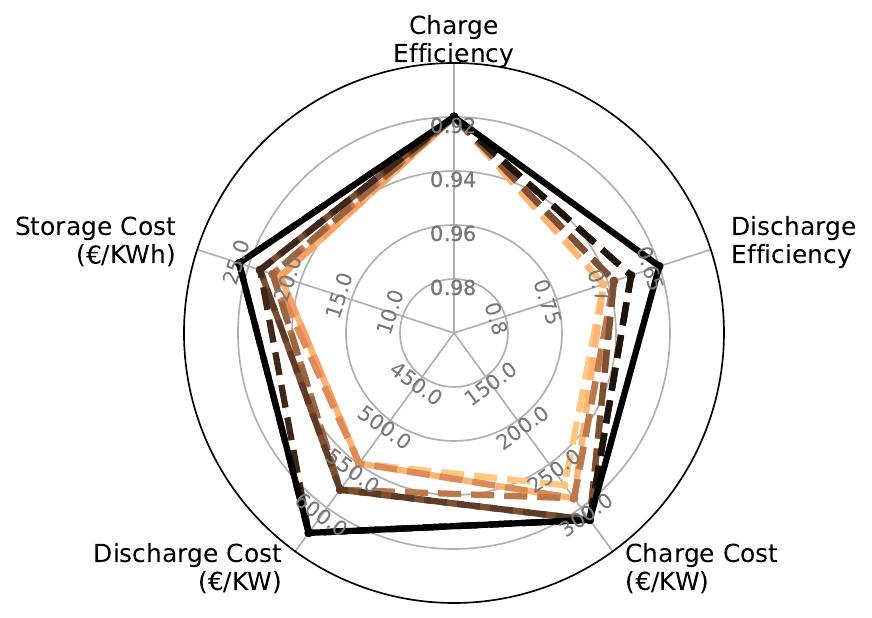}
    \caption{aCAES}
    \label{F:aCAES-Y2020-GCF0.85-alpha0.5-beta0.2}
  \end{subfigure}

  \begin{subfigure}[t]{0.7\textwidth}
    \centering
    \includegraphics[width=1.0\textwidth]{Figures/Denmark/Legend_2.pdf}
  \end{subfigure}

  \caption{Optimal development path of storage technologies \hspace{\textwidth}
  year = 2020, grid connection capacity factor $CF_{G_7}=0.85$, \hspace{\textwidth}
  slope parameter $\alpha=0.5$, improvement potential parameter $\beta=0.2$.}
  \label{F:Y2020-GCF0.85-alpha0.5-beta0.2}
\end{figure*}

\begin{figure*}
  \centering
  \begin{subfigure}[t]{0.49\textwidth}
    \centering
    \includegraphics[width=\textwidth]{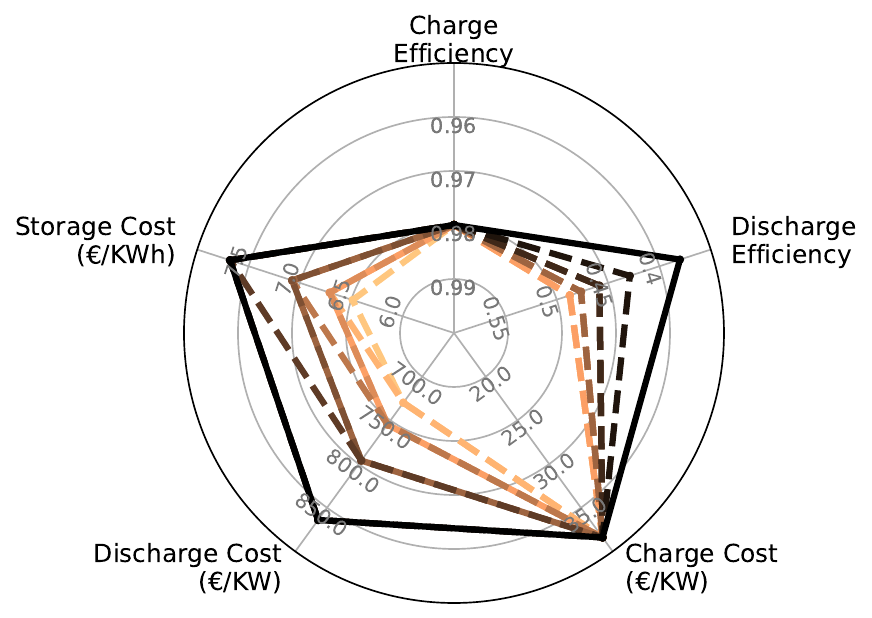}
    \caption{TES}
    \label{F:TES-Y2019-GCF0.85-alpha0.5-beta0.2-NoSolar}
  \end{subfigure}
  \hfill
  \begin{subfigure}[t]{0.49\textwidth}
    \centering
    \includegraphics[width=\textwidth]{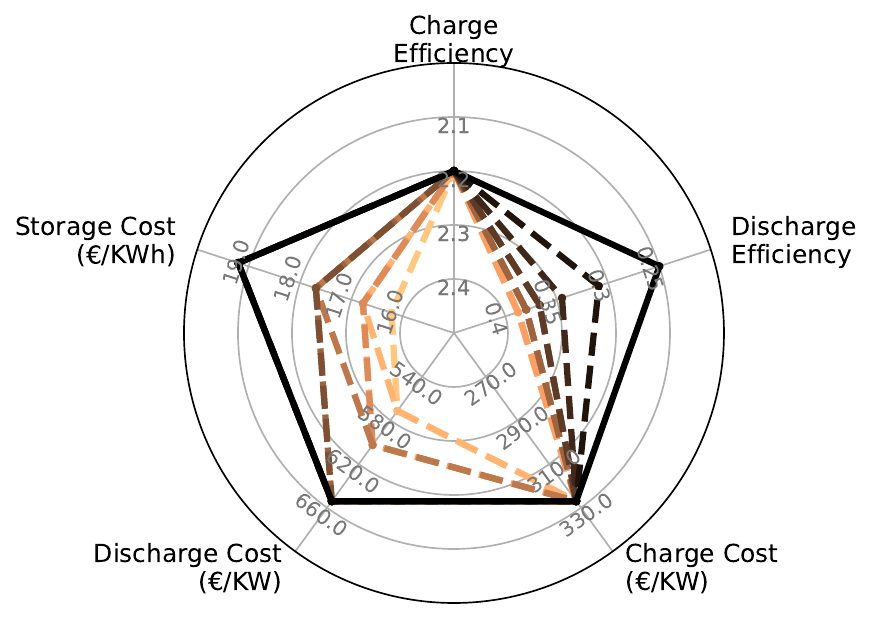}
    \caption{PTES}
    \label{F:PTES-Y2019-GCF0.85-alpha0.5-beta0.2-NoSolar}
  \end{subfigure}

  \vspace{0.5cm}

  \begin{subfigure}[t]{0.49\textwidth}
    \centering
    \includegraphics[width=\textwidth]{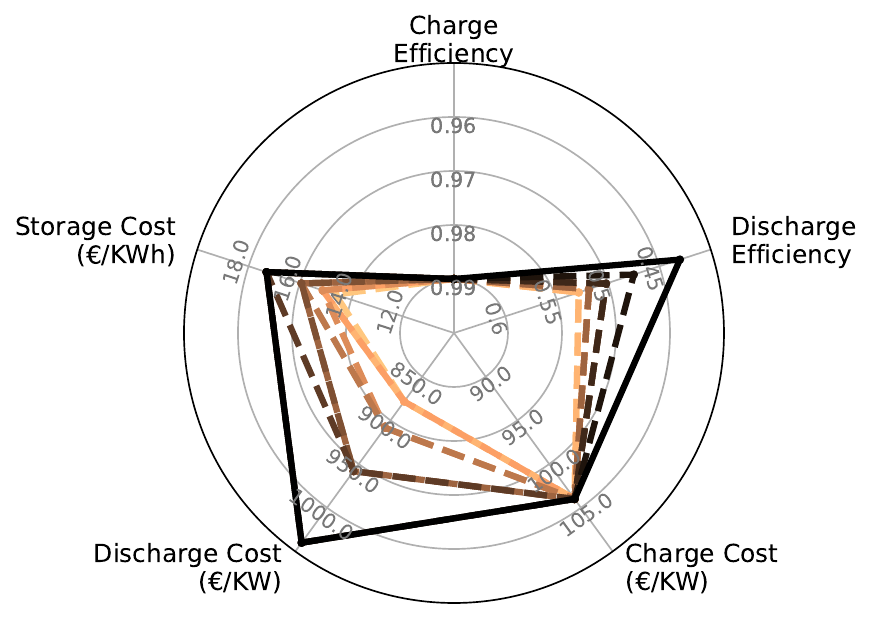}
    \caption{MSES}
    \label{F:MSES-Y2019-GCF0.85-alpha0.5-beta0.2-NoSolar}
  \end{subfigure}
  \hfill
  \begin{subfigure}[t]{0.49\textwidth}
    \centering
    \includegraphics[width=\textwidth]{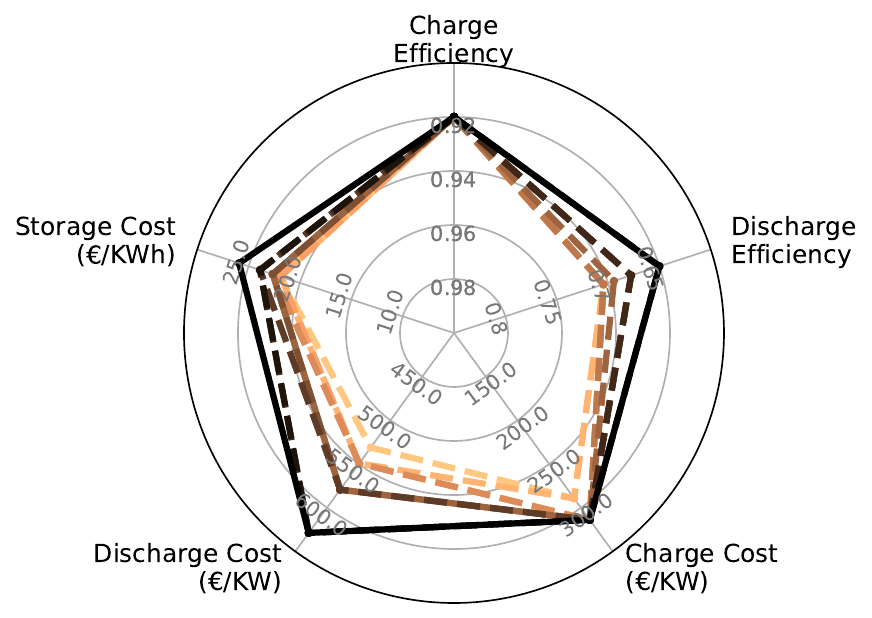}
    \caption{aCAES}
    \label{F:aCAES-Y2019-GCF0.85-alpha0.5-beta0.2-NoSolar}
  \end{subfigure}

  \begin{subfigure}[t]{0.7\textwidth}
    \centering
    \includegraphics[width=1.0\textwidth]{Figures/Denmark/Legend_2.pdf}
  \end{subfigure}

  \caption{Optimal development path of storage technologies \hspace{\textwidth}
  year = 2019, grid connection capacity factor $CF_{G_7}=0.85$, \hspace{\textwidth}
  slope parameter $\alpha=0.5$, improvement potential parameter $\beta=0.2$, no solar.}
  \label{F:Y2019-GCF0.85-alpha0.5-beta0.2-NoSolar}
\end{figure*}

\begin{figure*}
  \centering
  \begin{subfigure}[t]{0.49\textwidth}
    \centering
    \includegraphics[width=\textwidth]{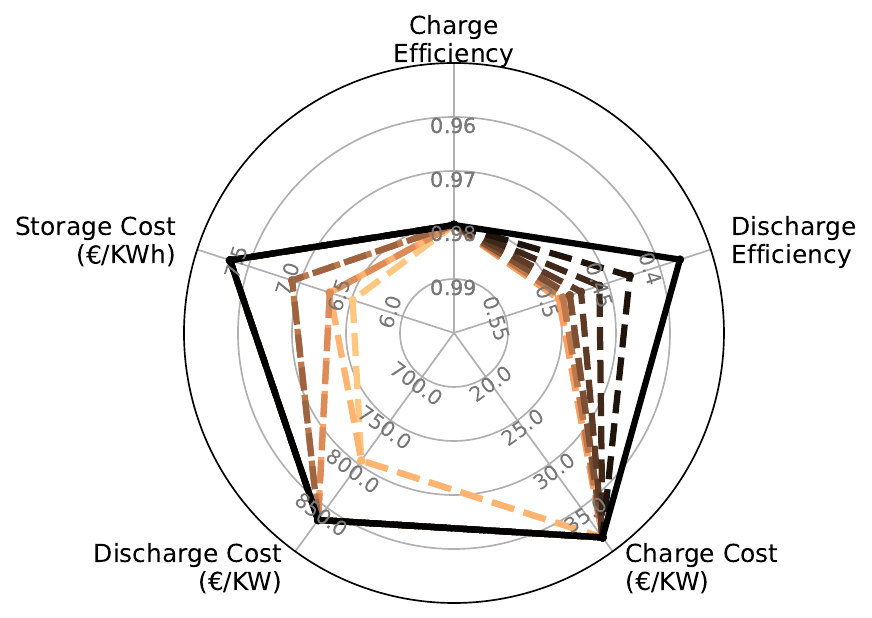}
    \caption{TES}
    \label{F:TES-Y2019-GCF0.85-alpha0.5-beta0.2-NoWind}
  \end{subfigure}
  \hfill
  \begin{subfigure}[t]{0.49\textwidth}
    \centering
    \includegraphics[width=\textwidth]{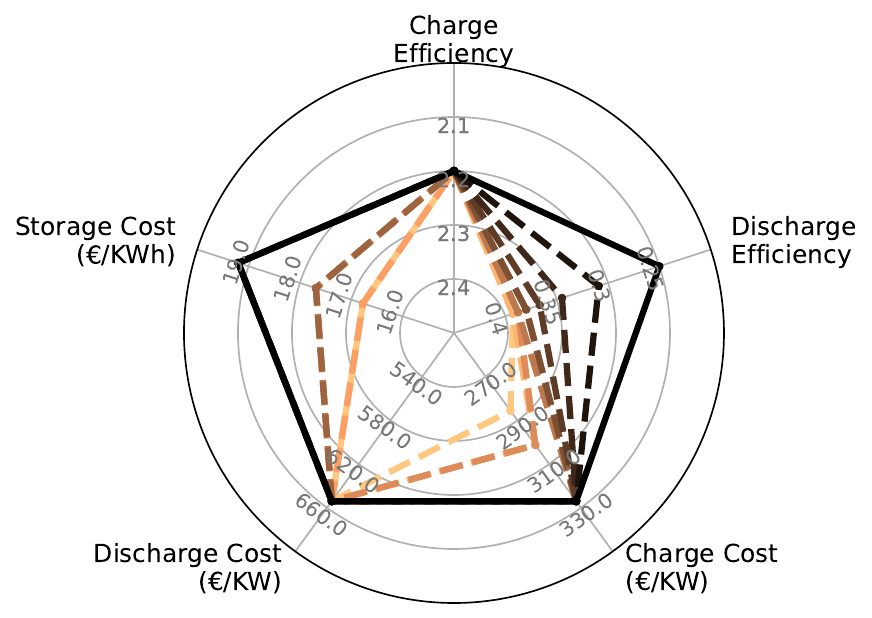}
    \caption{PTES}
    \label{F:PTES-Y2019-GCF0.85-alpha0.5-beta0.2-NoWind}
  \end{subfigure}

  \vspace{0.5cm}

  \begin{subfigure}[t]{0.49\textwidth}
    \centering
    \includegraphics[width=\textwidth]{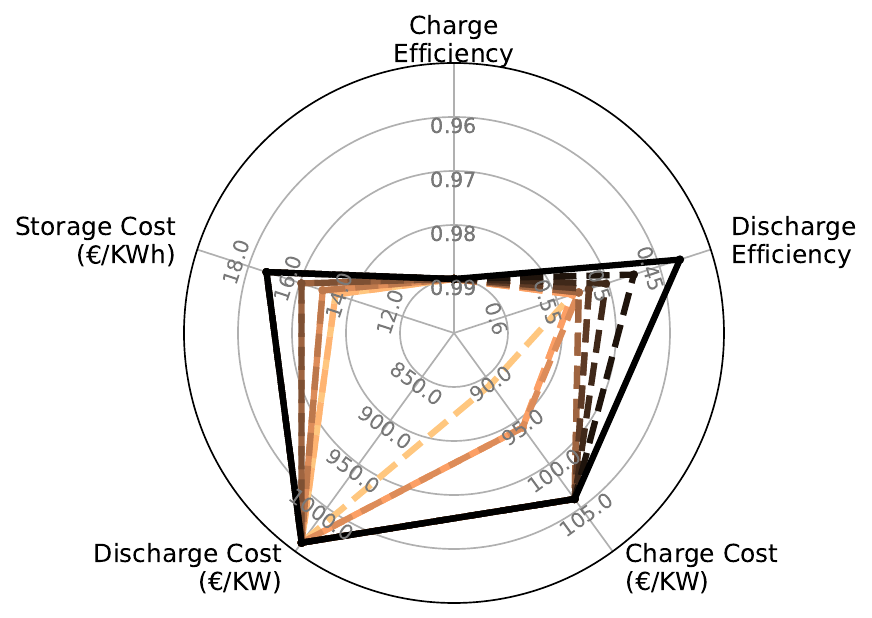}
    \caption{MSES}
    \label{F:MSES-Y2019-GCF0.85-alpha0.5-beta0.2-NoWind}
  \end{subfigure}
  \hfill
  \begin{subfigure}[t]{0.49\textwidth}
    \centering
    \includegraphics[width=\textwidth]{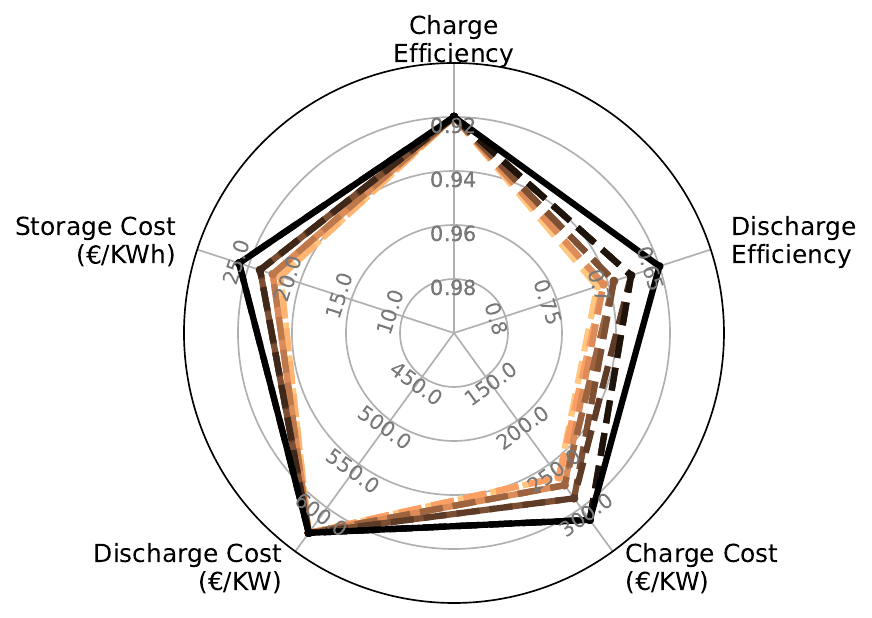}
    \caption{aCAES}
    \label{F:aCAES-Y2019-GCF0.85-alpha0.5-beta0.2-NoWind}
  \end{subfigure}

  \begin{subfigure}[t]{0.7\textwidth}
    \centering
    \includegraphics[width=1.0\textwidth]{Figures/Denmark/Legend_2.pdf}
  \end{subfigure}

  \caption{Optimal development path of storage technologies \hspace{\textwidth}
  year = 2019, grid connection capacity factor $CF_{G_7}=0.85$, \hspace{\textwidth}
  slope parameter $\alpha=0.5$, improvement potential parameter $\beta=0.2$, no wind.}
  \label{F:Y2019-GCF0.85-alpha0.5-beta0.2-NoWind}
\end{figure*}

Solar and wind capacity factors related to years 2018 (Fig.~\ref{F:Y2018-GCF0.85-alpha0.5-beta0.2}) and 2020 (Fig.~\ref{F:Y2020-GCF0.85-alpha0.5-beta0.2}) are considered. Two extreme scenarios whereby the renewable energy system operates without either solar (Fig.~\ref{F:Y2019-GCF0.85-alpha0.5-beta0.2-NoSolar}) or wind (Fig.~\ref{F:Y2019-GCF0.85-alpha0.5-beta0.2-NoWind}) energy sources are also considered. The extreme scenarios represent years or locations whereby solar or wind resources are significantly low.

\begin{figure*}
  \centering
  \begin{subfigure}[t]{0.49\textwidth}
    \centering
    \includegraphics[width=\textwidth]{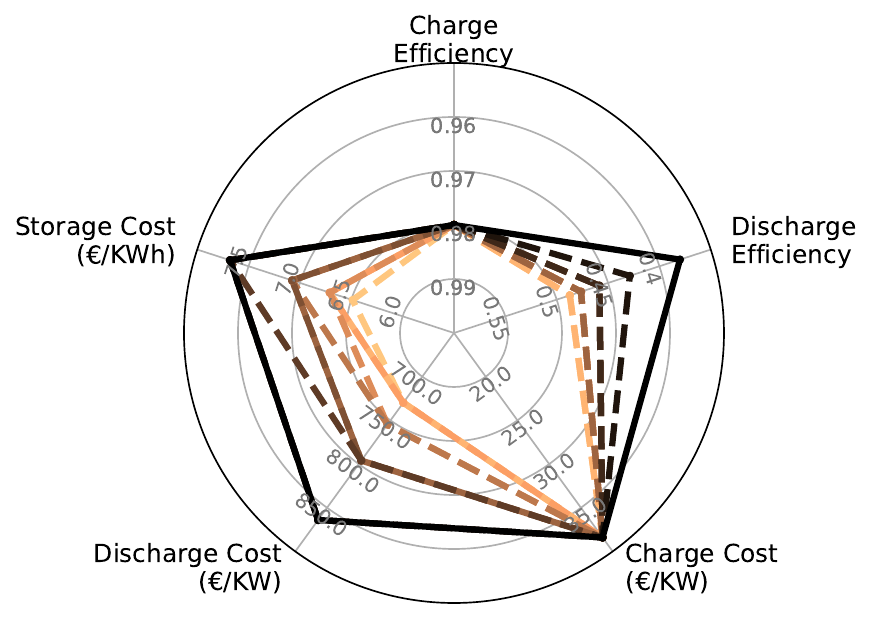}
    \caption{TES}
    \label{F:TES-Y2019-GCF0.75-alpha0.5-beta0.2}
  \end{subfigure}
  \hfill
  \begin{subfigure}[t]{0.49\textwidth}
    \centering
    \includegraphics[width=\textwidth]{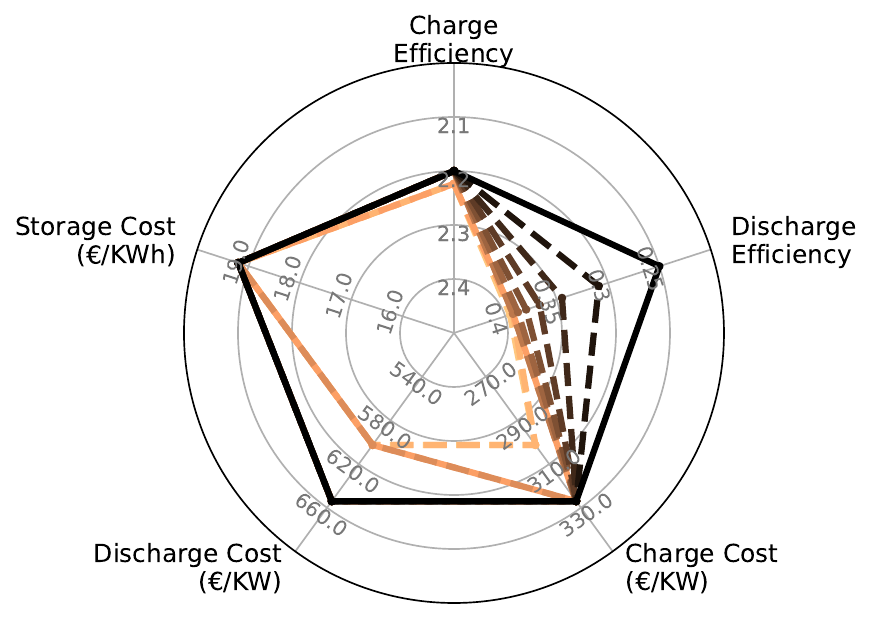}
    \caption{PTES}
    \label{F:PTES-Y2019-GCF0.75-alpha0.5-beta0.2}
  \end{subfigure}

  \vspace{0.5cm}

  \begin{subfigure}[t]{0.49\textwidth}
    \centering
    \includegraphics[width=\textwidth]{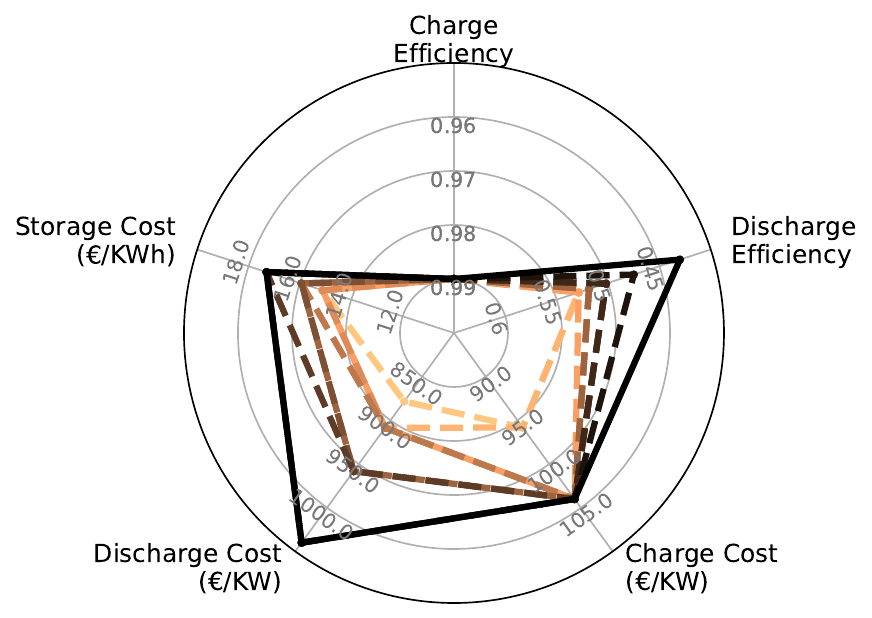}
    \caption{MSES}
    \label{F:MSES-Y2019-GCF0.75-alpha0.5-beta0.2}
  \end{subfigure}
  \hfill
  \begin{subfigure}[t]{0.49\textwidth}
    \centering
    \includegraphics[width=\textwidth]{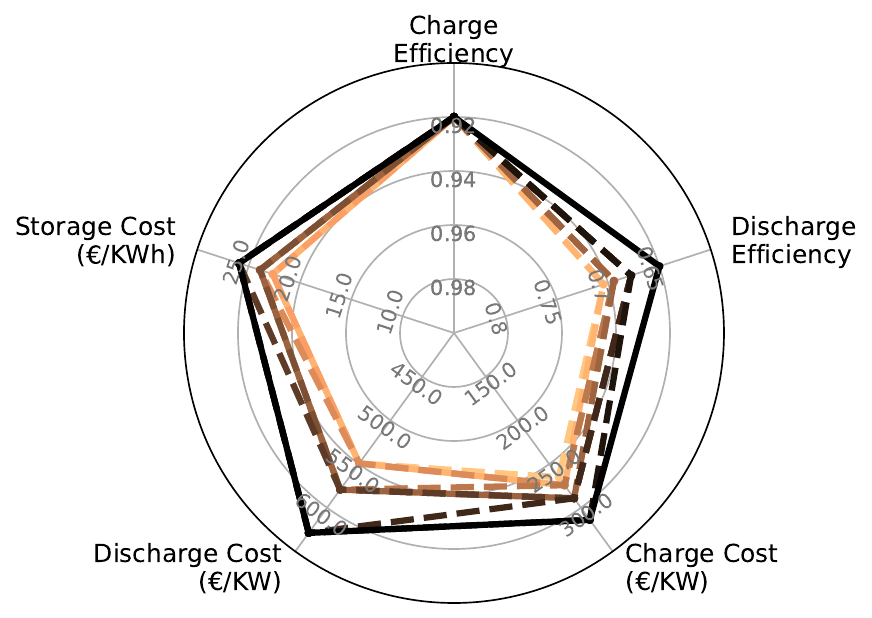}
    \caption{aCAES}
    \label{F:aCAES-Y2019-GCF0.75-alpha0.5-beta0.2}
  \end{subfigure}

  \begin{subfigure}[t]{0.7\textwidth}
    \centering
    \includegraphics[width=1.0\textwidth]{Figures/Denmark/Legend_2.pdf}
  \end{subfigure}

  \caption{Optimal development path of storage technologies \hspace{\textwidth}
  year = 2019, grid connection capacity factor $CF_{G_7}=0.75$, \hspace{\textwidth}
  slope parameter $\alpha=0.5$, improvement potential parameter $\beta=0.2$.}
  \label{F:Y2019-GCF0.75-alpha0.5-beta0.2}
\end{figure*}

\begin{figure*}
  \centering
  \begin{subfigure}[t]{0.49\textwidth}
    \centering
    \includegraphics[width=\textwidth]{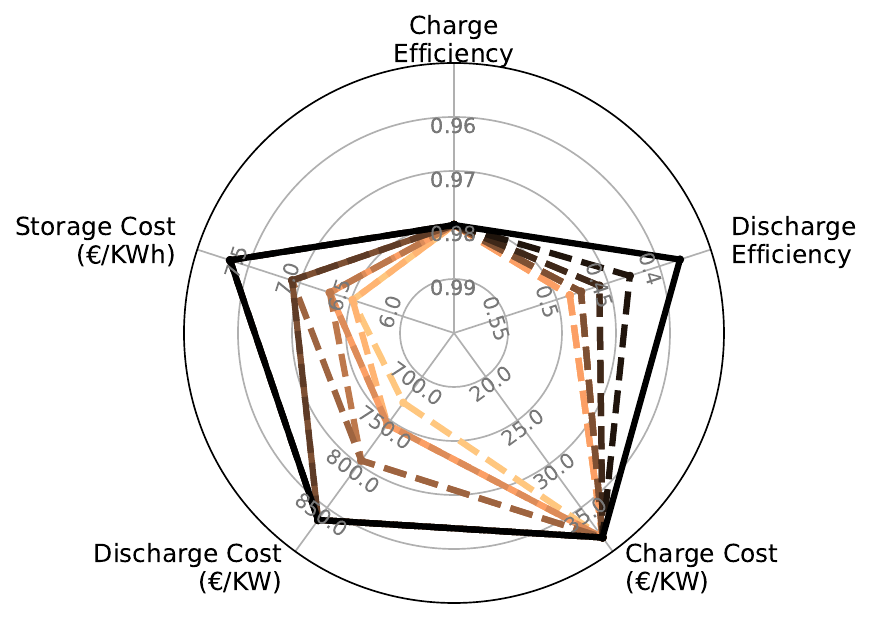}
    \caption{TES}
    \label{F:TES-Y2019-GCF0.95-alpha0.5-beta0.2}
  \end{subfigure}
  \hfill
  \begin{subfigure}[t]{0.49\textwidth}
    \centering
    \includegraphics[width=\textwidth]{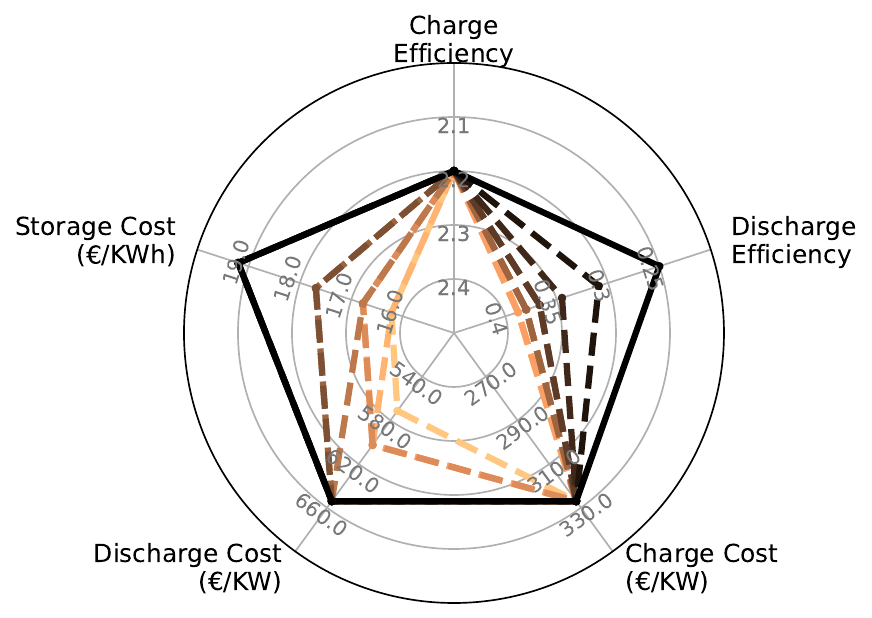}
    \caption{PTES}
    \label{F:PTES-Y2019-GCF0.95-alpha0.5-beta0.2}
  \end{subfigure}

  \vspace{0.5cm}

  \begin{subfigure}[t]{0.49\textwidth}
    \centering
    \includegraphics[width=\textwidth]{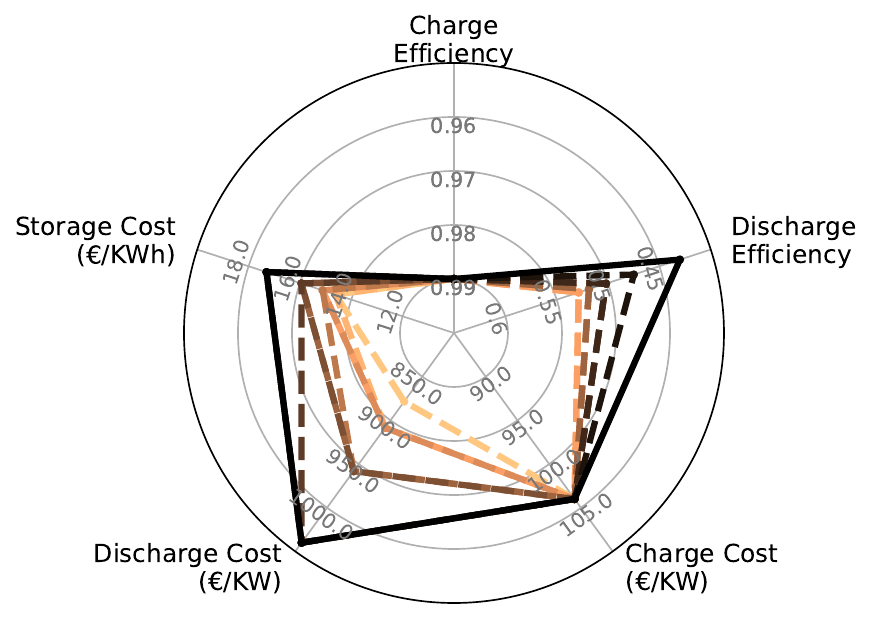}
    \caption{MSES}
    \label{F:MSES-Y2019-GCF0.95-alpha0.5-beta0.2}
  \end{subfigure}
  \hfill
  \begin{subfigure}[t]{0.49\textwidth}
    \centering
    \includegraphics[width=\textwidth]{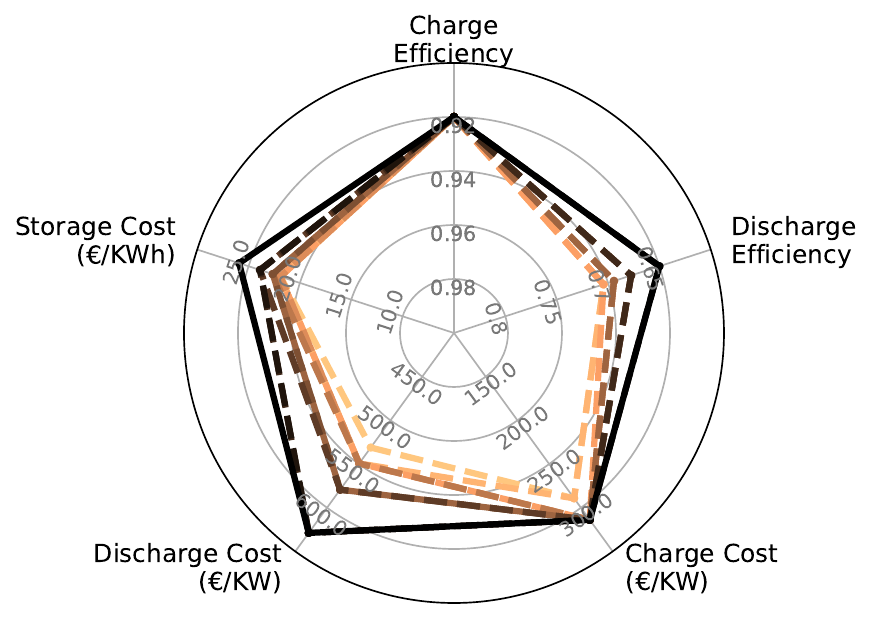}
    \caption{aCAES}
    \label{F:aCAES-Y2019-GCF0.95-alpha0.5-beta0.2}
  \end{subfigure}

  \begin{subfigure}[t]{0.7\textwidth}
    \centering
    \includegraphics[width=1.0\textwidth]{Figures/Denmark/Legend_2.pdf}
  \end{subfigure}

  \caption{Optimal development path of storage technologies \hspace{\textwidth}
  year = 2019, grid connection capacity factor $CF_{G_7}=0.95$, \hspace{\textwidth}
  slope parameter $\alpha=0.5$, improvement potential parameter $\beta=0.2$.}
  \label{F:Y2019-GCF0.95-alpha0.5-beta0.2}
\end{figure*}

Grid connection capacity factors equal to 0.75 (Fig.~\ref{F:Y2019-GCF0.75-alpha0.5-beta0.2}) and 0.95 (Fig.~\ref{F:Y2019-GCF0.95-alpha0.5-beta0.2}) are considered. The lower capacity factor facilitates assessing the impact on the storage parameters whereby the significance of energy storage in the hybrid energy system is relatively low. In contrast, the higher capacity factor enables assessing the impact on the storage parameters whereby energy storage is essential in the hybrid energy system.

\begin{figure*}
  \centering
  \begin{subfigure}[t]{0.49\textwidth}
    \centering
    \includegraphics[width=\textwidth]{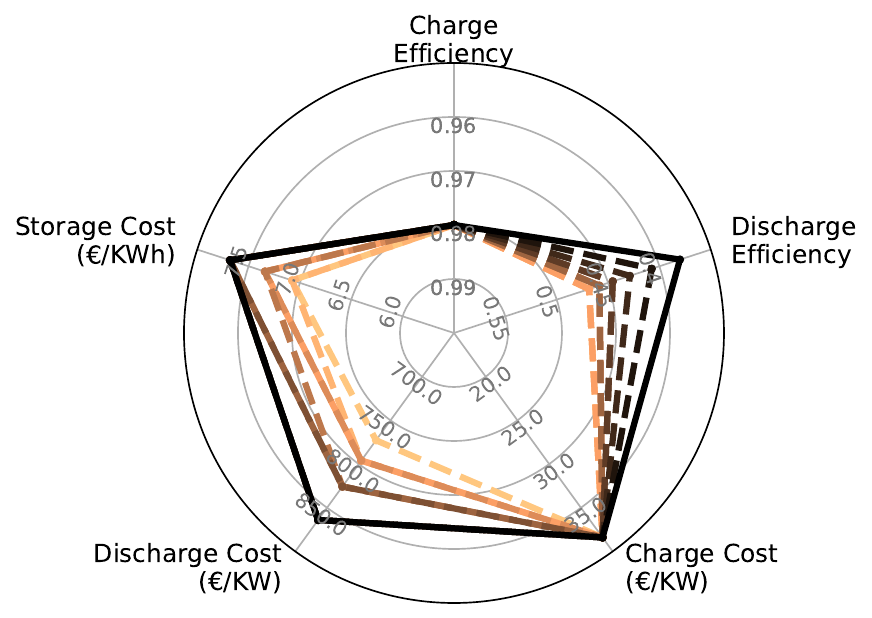}
    \caption{TES}
    \label{F:TES-Y2019-GCF0.85-alpha0.25-beta0.2}
  \end{subfigure}
  \hfill
  \begin{subfigure}[t]{0.49\textwidth}
    \centering
    \includegraphics[width=\textwidth]{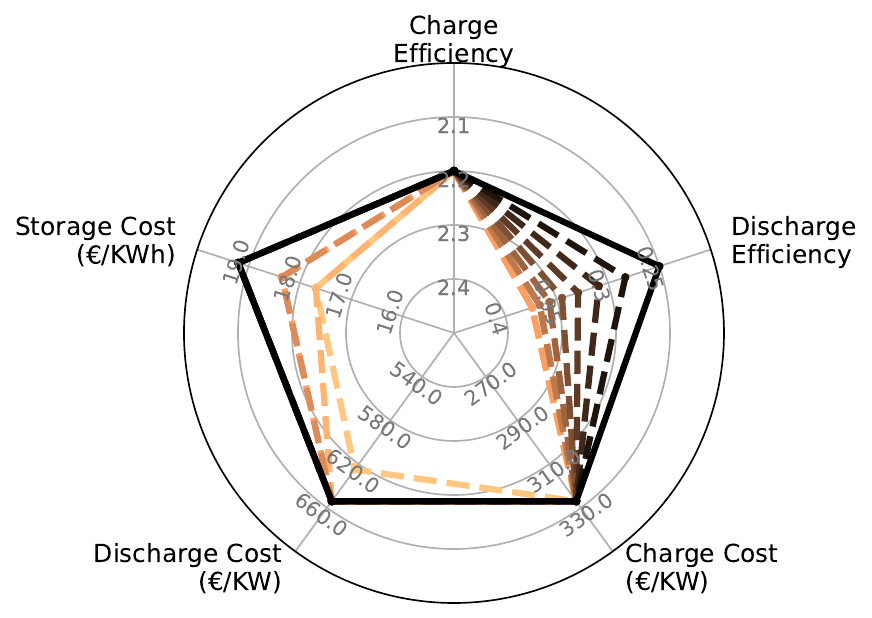}
    \caption{PTES}
    \label{F:PTES-Y2019-GCF0.85-alpha0.25-beta0.2}
  \end{subfigure}

  \vspace{0.5cm}

  \begin{subfigure}[t]{0.49\textwidth}
    \centering
    \includegraphics[width=\textwidth]{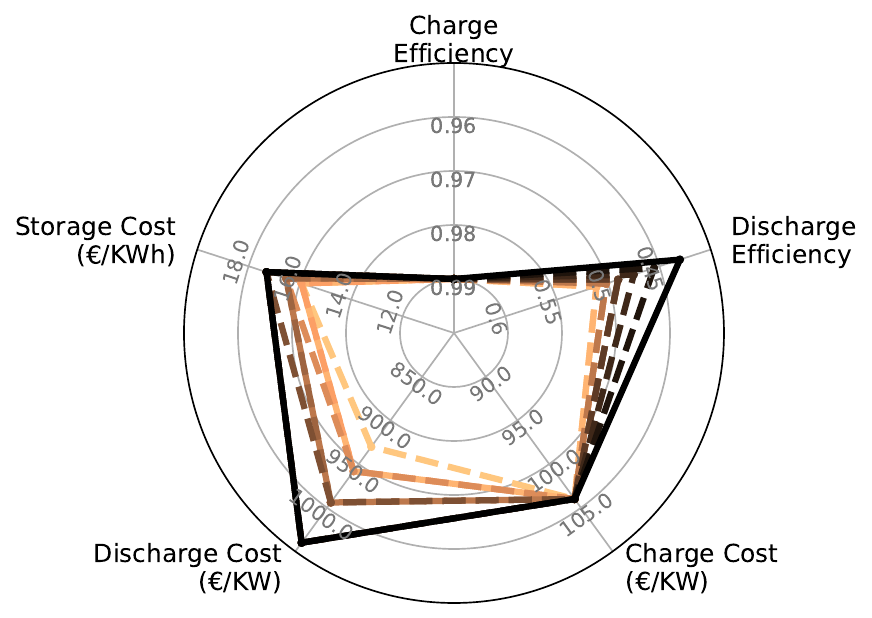}
    \caption{MSES}
    \label{F:MSES-Y2019-GCF0.85-alpha0.25-beta0.2}
  \end{subfigure}
  \hfill
  \begin{subfigure}[t]{0.49\textwidth}
    \centering
    \includegraphics[width=\textwidth]{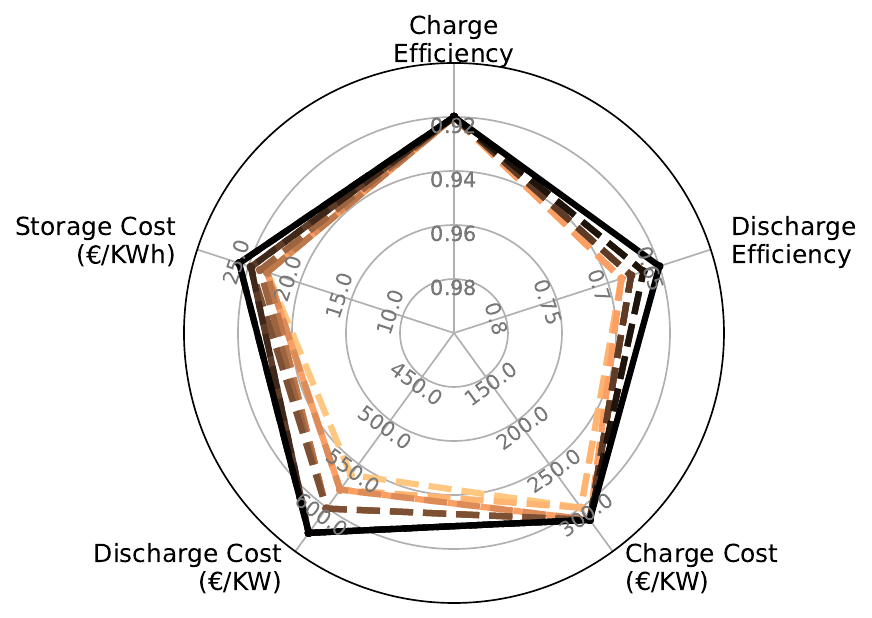}
    \caption{aCAES}
    \label{F:aCAES-Y2019-GCF0.85-alpha0.25-beta0.2}
  \end{subfigure}

  \begin{subfigure}[t]{0.7\textwidth}
    \centering
    \includegraphics[width=1.0\textwidth]{Figures/Denmark/Legend_2.pdf}
  \end{subfigure}

  \caption{Optimal development path of storage technologies \hspace{\textwidth}
  year = 2019, grid connection capacity factor $CF_{G_7}=0.85$, \hspace{\textwidth}
  slope parameter $\alpha=0.25$, improvement potential parameter $\beta=0.2$.}
  \label{F:Y2019-GCF0.85-alpha0.25-beta0.2}
\end{figure*}

\begin{figure*}
  \centering
  \begin{subfigure}[t]{0.49\textwidth}
    \centering
    \includegraphics[width=\textwidth]{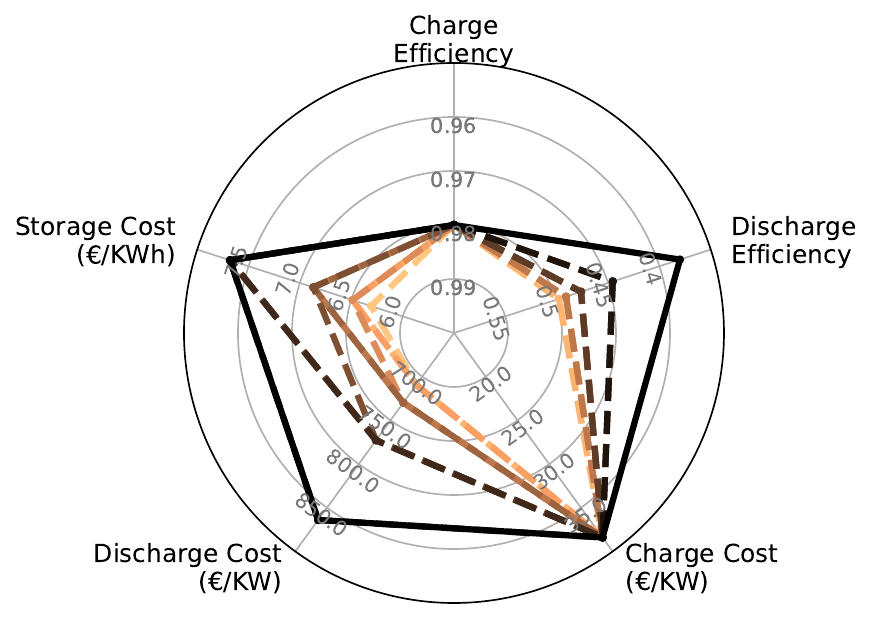}
    \caption{TES}
    \label{F:TES-Y2019-GCF0.85-alpha0.75-beta0.2}
  \end{subfigure}
  \hfill
  \begin{subfigure}[t]{0.49\textwidth}
    \centering
    \includegraphics[width=\textwidth]{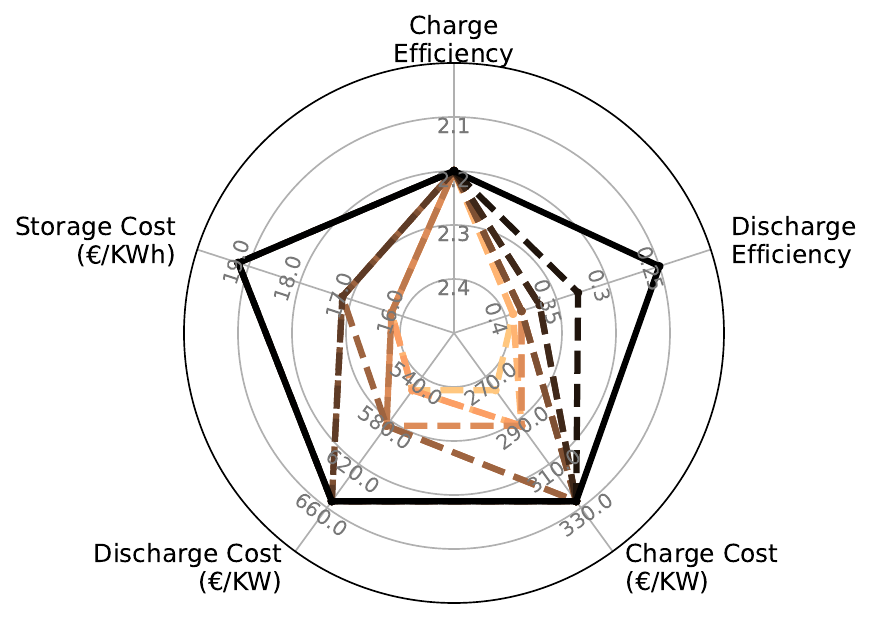}
    \caption{PTES}
    \label{F:PTES-Y2019-GCF0.85-alpha0.75-beta0.2}
  \end{subfigure}

  \vspace{0.5cm}

  \begin{subfigure}[t]{0.49\textwidth}
    \centering
    \includegraphics[width=\textwidth]{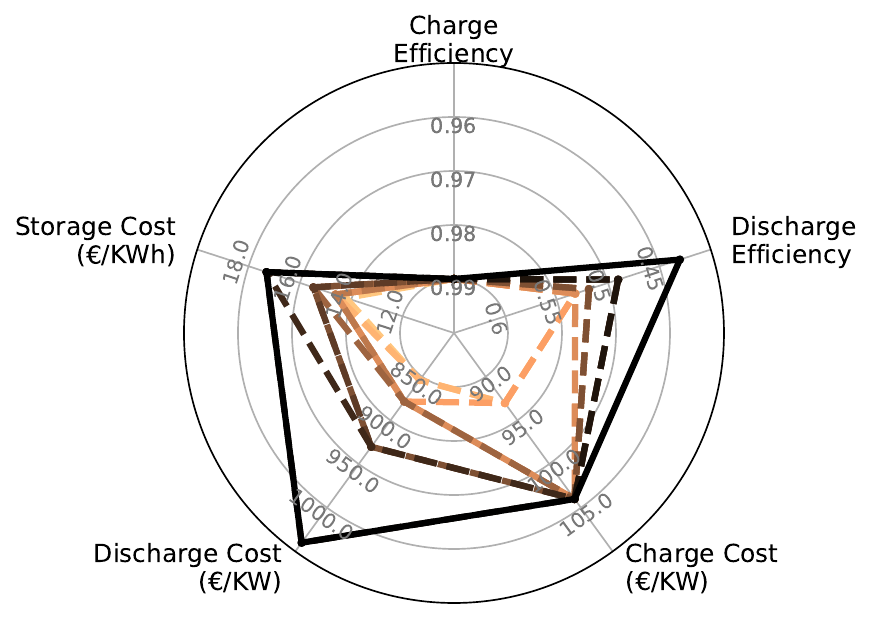}
    \caption{MSES}
    \label{F:MSES-Y2019-GCF0.85-alpha0.75-beta0.2}
  \end{subfigure}
  \hfill
  \begin{subfigure}[t]{0.49\textwidth}
    \centering
    \includegraphics[width=\textwidth]{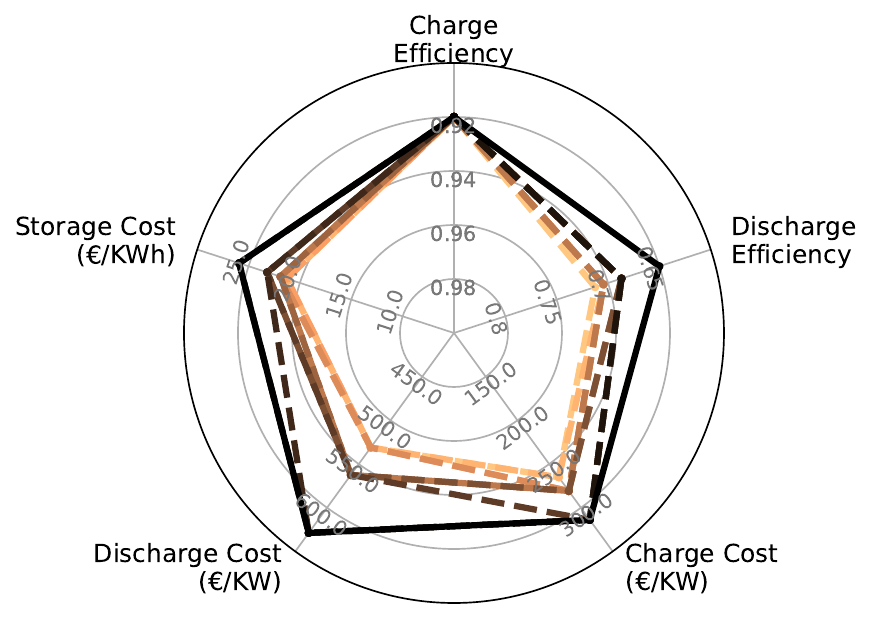}
    \caption{aCAES}
    \label{F:aCAES-Y2019-GCF0.85-alpha0.75-beta0.2}
  \end{subfigure}

  \begin{subfigure}[t]{0.7\textwidth}
    \centering
    \includegraphics[width=1.0\textwidth]{Figures/Denmark/Legend_2.pdf}
  \end{subfigure}

  \caption{Optimal development path of storage technologies \hspace{\textwidth}
  year = 2019, grid connection capacity factor $CF_{G_7}=0.85$, \hspace{\textwidth}
  slope parameter $\alpha=0.75$, improvement potential parameter $\beta=0.2$.}
  \label{F:Y2019-GCF0.85-alpha0.75-beta0.2}
\end{figure*}

\begin{figure*}
  \centering
  \begin{subfigure}[t]{0.49\textwidth}
    \centering
    \includegraphics[width=\textwidth]{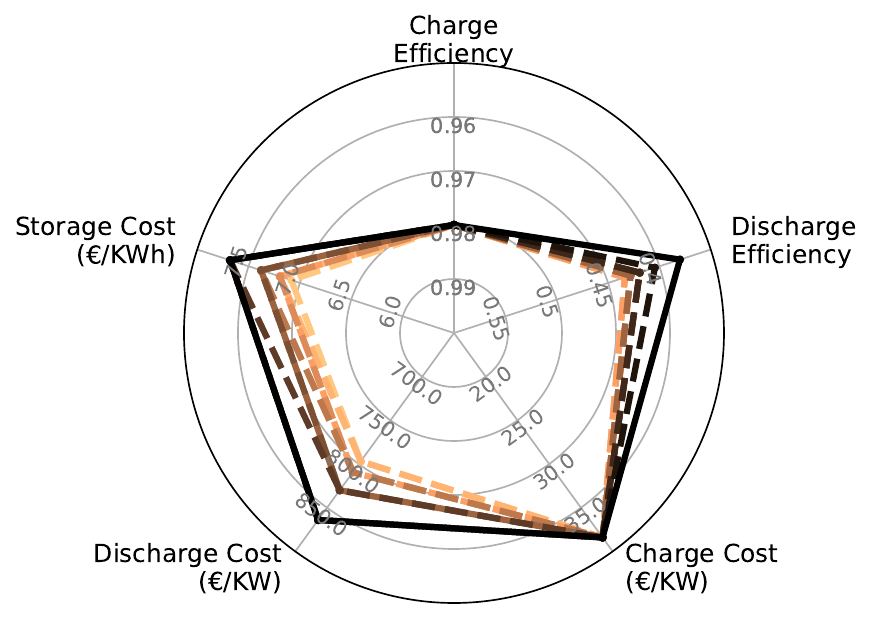}
    \caption{TES}
    \label{F:TES-Y2019-GCF0.85-alpha0.5-beta0.1}
  \end{subfigure}
  \hfill
  \begin{subfigure}[t]{0.49\textwidth}
    \centering
    \includegraphics[width=\textwidth]{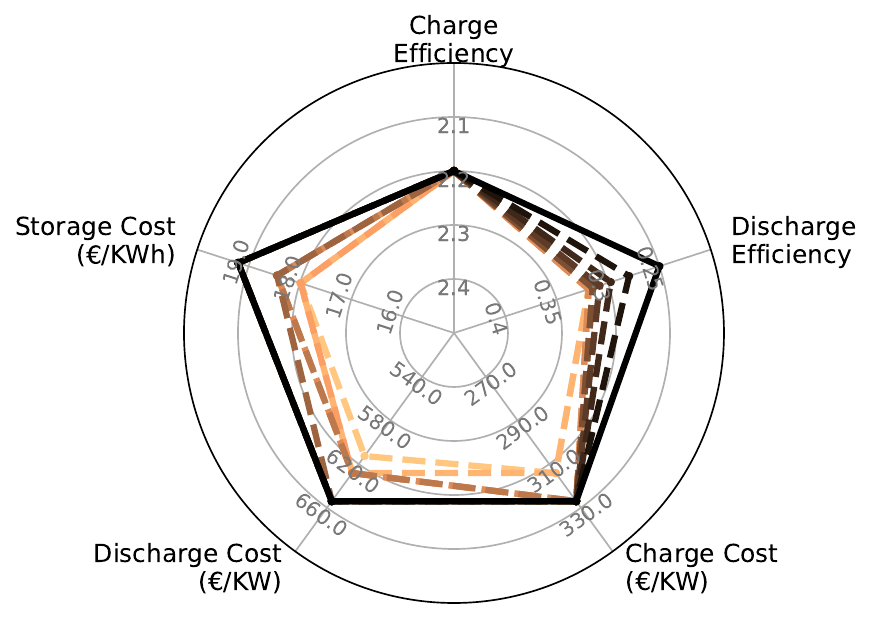}
    \caption{PTES}
    \label{F:PTES-Y2019-GCF0.85-alpha0.5-beta0.1}
  \end{subfigure}

  \vspace{0.5cm}

  \begin{subfigure}[t]{0.49\textwidth}
    \centering
    \includegraphics[width=\textwidth]{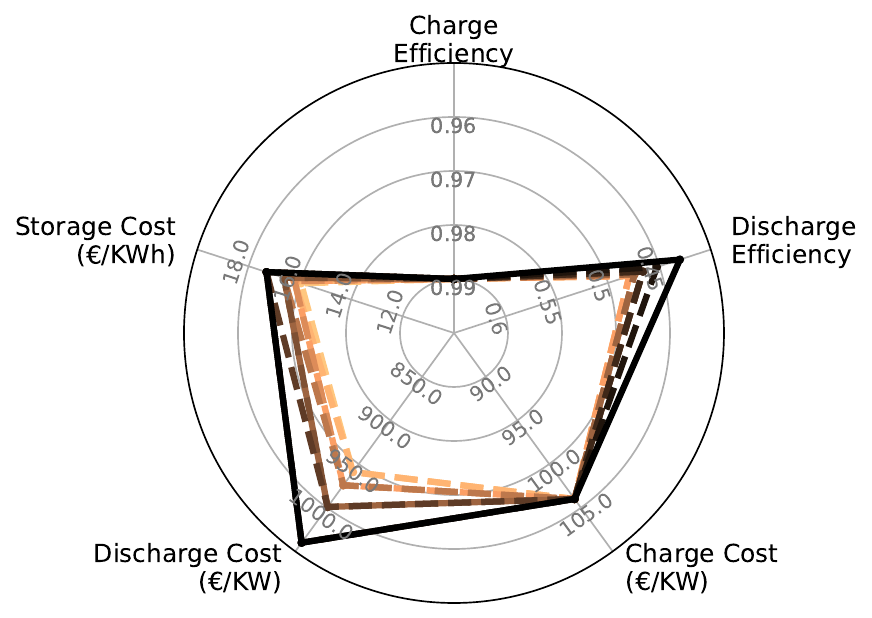}
    \caption{MSES}
    \label{F:MSES-Y2019-GCF0.85-alpha0.5-beta0.1}
  \end{subfigure}
  \hfill
  \begin{subfigure}[t]{0.49\textwidth}
    \centering
    \includegraphics[width=\textwidth]{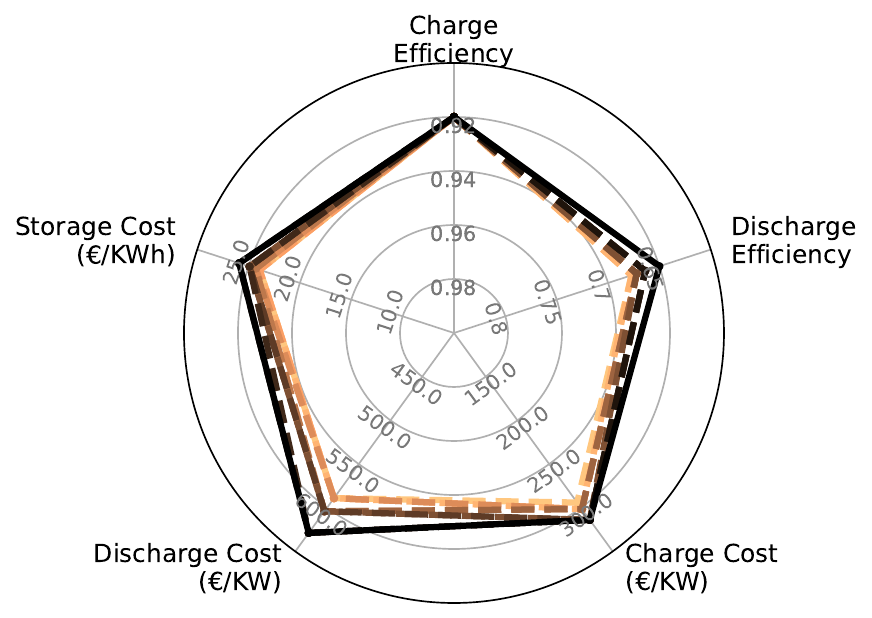}
    \caption{aCAES}
    \label{F:aCAES-Y2019-GCF0.85-alpha0.5-beta0.1}
  \end{subfigure}

  \begin{subfigure}[t]{0.7\textwidth}
    \centering
    \includegraphics[width=1.0\textwidth]{Figures/Denmark/Legend_2.pdf}
  \end{subfigure}

  \caption{Optimal development path of storage technologies \hspace{\textwidth}
  year = 2019, grid connection parameter $CF_{G_7}=0.85$, \hspace{\textwidth}
  slope parameter $\alpha=0.5$, improvement potential parameter $\beta=0.1$.}
  \label{F:Y2019-GCF0.85-alpha0.5-beta0.1}
\end{figure*}

\begin{figure*}
  \centering
  \begin{subfigure}[t]{0.49\textwidth}
    \centering
    \includegraphics[width=\textwidth]{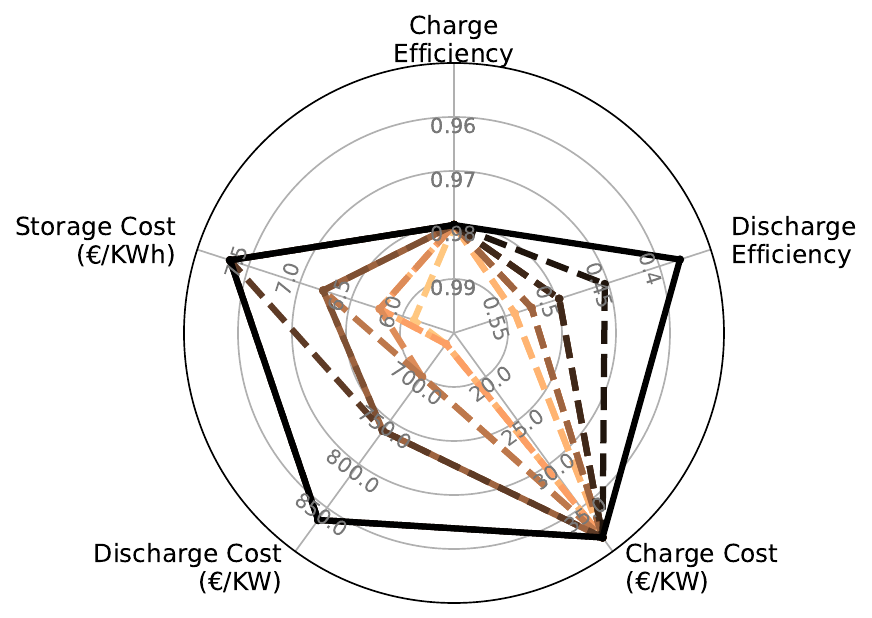}
    \caption{TES}
    \label{F:TES-Y2019-GCF0.85-alpha0.5-beta0.3}
  \end{subfigure}
  \hfill
  \begin{subfigure}[t]{0.49\textwidth}
    \centering
    \includegraphics[width=\textwidth]{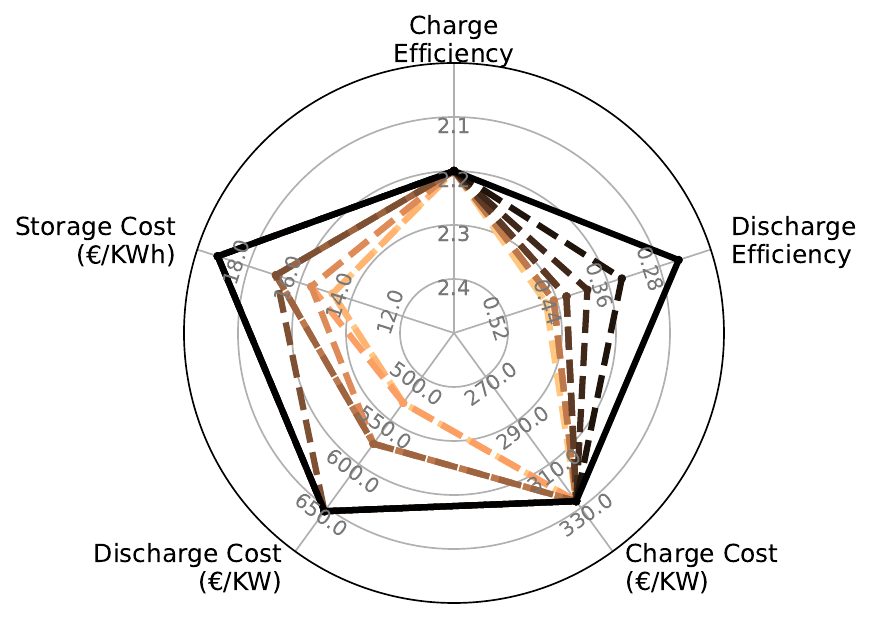}
    \caption{PTES}
    \label{F:PTES-Y2019-GCF0.85-alpha0.5-beta0.3}
  \end{subfigure}

  \vspace{0.5cm}

  \begin{subfigure}[t]{0.49\textwidth}
    \centering
    \includegraphics[width=\textwidth]{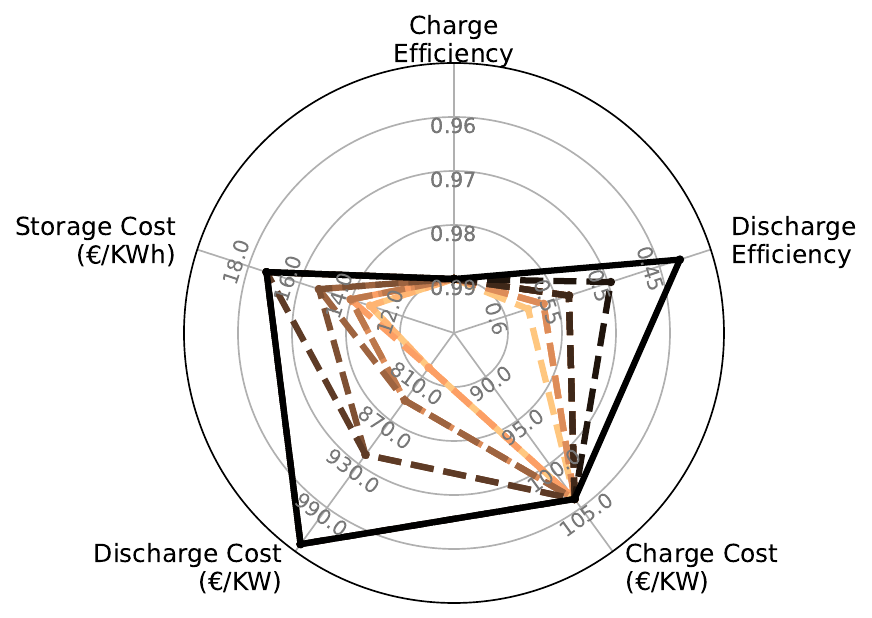}
    \caption{MSES}
    \label{F:MSES-Y2019-GCF0.85-alpha0.5-beta0.3}
  \end{subfigure}
  \hfill
  \begin{subfigure}[t]{0.49\textwidth}
    \centering
    \includegraphics[width=\textwidth]{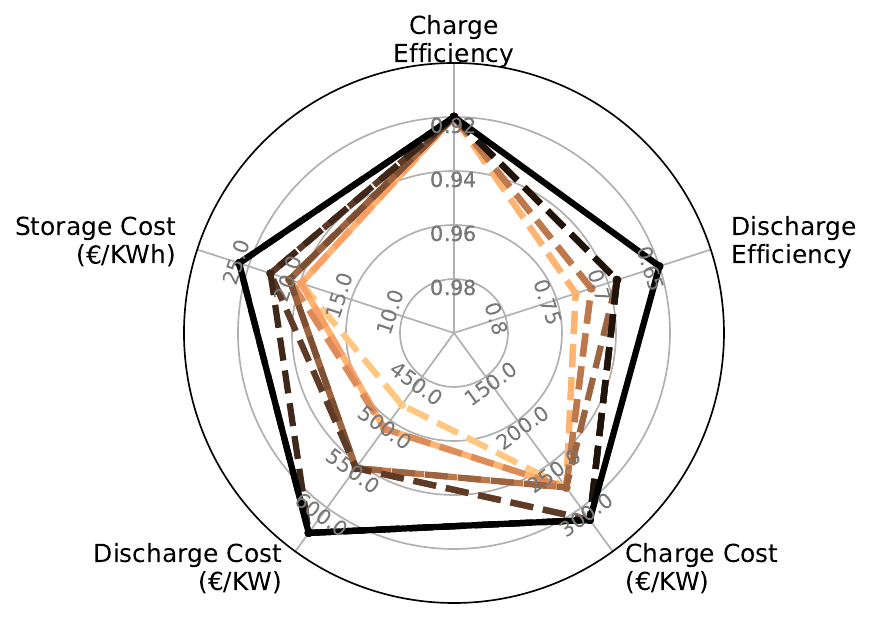}
    \caption{aCAES}
    \label{F:aCAES-Y2019-GCF0.85-alpha0.5-beta0.3}
  \end{subfigure}

  \begin{subfigure}[t]{0.7\textwidth}
    \centering
    \includegraphics[width=1.0\textwidth]{Figures/Denmark/Legend_2.pdf}
  \end{subfigure}

  \caption{Optimal development path of storage technologies \hspace{\textwidth}
  year = 2019, grid connection capacity factor $CF_{G_7}=0.85$, \hspace{\textwidth}
  slope parameter $\alpha=0.5$, improvement potential parameter $\beta=0.3$.}
  \label{F:Y2019-GCF0.85-alpha0.5-beta0.3}
\end{figure*}

The values related to the investment optimisation parameters are $\alpha$ equal 0.25 (Fig.~\ref{F:Y2019-GCF0.85-alpha0.25-beta0.2}) and 0.75 (Fig.~\ref{F:Y2019-GCF0.85-alpha0.75-beta0.2}), and $\beta$ equal 0.1 (Fig.~\ref{F:Y2019-GCF0.85-alpha0.5-beta0.1}) and 0.3 (Fig.~\ref{F:Y2019-GCF0.85-alpha0.5-beta0.3}). The lower $\alpha$ value represents a scenario whereby a unit investment has a relatively small effect on improving the storage parameters. In contrast, the higher $\alpha$ value represents a scenario whereby a unit investment has a relatively large effect on improving the storage parameters. The lower $\beta$ value represents a relatively mature technology, whereas a higher $\beta$ value represents a relatively less mature technology with higher potential for improvements.

\nocite{*}

\bibliography{Bibliography}

\end{document}